
\input epsf
\magnification\magstep1



\hbadness=10000      
\vbadness=10000  

\font\eightit=cmti8  
\font\eightrm=cmr8 \font\eighti=cmmi8                 
\font\eightsy=cmsy8 
           \font\sixrm=cmr6



\def\eightpoint{\normalbaselineskip=10pt 
\def\rm{\eightrm\fam0} \let\it\eightit
\textfont0=\eightrm \scriptfont0=\sixrm 
\textfont1=\eighti \scriptfont1=\seveni
\textfont2=\eightsy \scriptfont2=\sevensy 
\normalbaselines \eightrm
\parindent=1em}



\def\eq#1{{\noexpand\rm(#1)}}          
\newcount\eqcounter                    
\eqcounter=0                           
\def\numeq{\global\advance\eqcounter by 1\eq{\the\eqcounter}}           
\def\relativeq#1{{\advance\eqcounter by #1\eq{\the\eqcounter}}}


\def\namelasteq#1{\global\edef#1{{\eq{\the\eqcounter}}}}  

\def\TR{T_R}		
\def\cteI{\Bigl({1\over(4\pi)^2\varepsilon}\Bigr)}

\def\cite#1{{\rm[#1]}}                 
\def\Trace{{\rm Tr}\,}

\def\ren{{\rm ren}}


\newif\ifstartsec                   

\outer\def\section#1{\vskip 0pt plus .15\vsize \penalty -250
\vskip 0pt plus -.15\vsize \bigskip \startsectrue
\message{#1}\centerline{\bf#1}\nobreak\noindent}

\def\subsection#1{\ifstartsec\medskip\else\bigskip\fi \startsecfalse
\noindent{\it#1}\penalty100\medskip}

\def\refno#1. #2\par{\smallskip\item{\rm\lbrack#1\rbrack}#2\par}

\hyphenation{geo-me-try}

\bigskip


\def\Tony{1}
\def\Makeenko{2}
\def\Doplicher{3}
\def\CDS{4}
\def\Mtheory{5}
\def\Branes{6}
\def\Filk{7}
\def\Continuum{8}
\def\Lattice{9}
\def\Phenomenon{10}
\def\MRS{11}
\def\UVIR{12}
\def\NCBPHZ{13}
\def\BPHZ{14}
\def\Susy{15}
\def\Wilson{16}
\def\Firstorder{17}
\def\Nonren{18}
\def\QAP{19}
\def\Algren{20}
\def\Unitarity{21}
\def\Books{22}
\def\Hooft{23}
\def\BM{24}
\def\NCUN{25}


\rightline{FT/UCM--60--2000}

\vskip 1cm

\centerline{\bf The BRS invariance of noncommutative $U(N)$ Yang-Mills theory}
\centerline{\bf  at the one-loop level}

\bigskip

\centerline{\rm C. P. Mart{\'\i}n* and D. S\'anchez-Ruiz\dag}
\medskip
\centerline{\eightit Departamento de F{\'\i}sica Te\'orica I,						Universidad Complutense, 28040 Madrid, Spain}
\vfootnote*{email: {\tt carmelo@elbereth.fis.ucm.es}}
\vfootnote\dag{email: {\tt domingos@eucmos.sim.ucm.es}}

\bigskip\bigskip

\begingroup\narrower\narrower
\eightpoint
We show that $U(N)$ Yang-Mills theory on noncommutative Minkowski space-time
can be renormalized, in a BRS invariant way, at the one-loop level, by 
multiplicative dimensional renormalization of its coupling constant, 
its gauge parameter and its fields. It is shown that the Slavnov-Taylor 
equation, the gauge-fixing equation and the ghost equation hold, up to 
order $\hbar$, for the MS renormalized noncommutative $U(N)$ Yang-Mills theory.
We give the value of the pole part of every 1PI diagram which is UV divergent.

\par
\endgroup 
\vskip 2cm

\section{1.- Introduction}

Noncommutative field theories occur both in the ordinary  
(commutative space-time) field  theory setting and in the realm  of 
string theory. The study of the large N limit of ordinary field theories 
naturally leads to field theories over noncommutative 
space~\cite{\Tony, \Makeenko}.
General relativity and Heisenberg's uncertainty principles 
give rise, when strong gravitational fields are on, to space-times defined by 
noncommuting operators~\cite{\Doplicher}, whereupon it arises the
need to define quantum field theories over noncommutative space-times.
Super Yang-Mills theories on   noncommutative tori
occur in compactifications of M(atrix)-theory~\cite{\CDS},  M(atrix)-theory  
on noncommutative tori being the subject of a good many 
papers~\cite{\Mtheory}. Theories of strings ending on D-branes in 
the presence of a NS-NS B-field lead to noncommutative space-times; their 
infinite tension limit being --if unitarity allows it-- certain 
noncommutative field theories~\cite{\Branes}. It is therefore no wonder that
a sizeable amount of work has been put in understanding, either in the 
continuum~\cite{\Filk, \Continuum} or on the lattice~\cite{\Lattice}, whether
quantum field theories make sense on noncommutative space-times. 
Applications to collider physics and Cosmology have just begun to come up~\cite{\Phenomenon}.

 Quantum field theories on noncommutative space-time present a characteristic
connection between UV and IR scales: the virtual high-momenta modes 
contributing to a given Green function yield, when moving around a planar 
loop, an UV divergence, but  give rise to an IR divergence -even if the 
classical Lagrangian has only massive fields- as they propagate 
along  a nonplanar loop. This is the UV/IR mixing unveiled in 
ref.~\cite{\MRS}, which has been further investigated in refs.~\cite{\UVIR}.
The new --as regards to quantum field theory on commutative space-time-- 
IR divergences that occur in noncommutative field theories makes it impossible
~\cite{\NCBPHZ}, beyond a few loops, that these theories  be renormalizable 
{\it \'a la} Bologiubov-Parasiuk-Hepp-Zimmerman~\cite{\BPHZ}, if supersymmetry 
is not called in~\cite{\Susy}. Besides, 
they lack a Wilsonian action~\cite{\MRS}, which  puts in jeopardy 
the implementation in noncommutative field  theories of Wilson's 
renormalization group program~\cite{\Wilson}; and,  hence, the existence 
of a  continuum limit for these theories. The existence of a continuum 
limit for noncommutative field theories has been studied in 
ref.~\cite{\Firstorder}. 

It is well known~\cite{\Filk} that, at the one-loop level, only if a 
diagram is planar it can be UV divergent and that the momentum structure 
of this divergence, should it exist, is the product of a  polynomial  
of the appropriate dimension of the external momenta 
(the UV degree of divergence of the Feynman loop integral) by suitable Moyal 
phases.  Besides, if the noncommutative diagram has an ordinary counterpart 
(the diagram when space-time is commutative), the polynomial of the external 
momenta which carries the UV divergence is the same for both diagrams.
This might lead us to think  that noncommutative field theories are 
always one-loop renormalizable,  if their ordinary counterpart
is; which would in turn render almost trivial the issue of the one-loop 
renormalizability of noncommutative field theories. One cannot be 
more mistaken. There are certain  $*$-deformations of $\lambda \phi^4$ 
that are shown not to be renormalizable at the one-loop 
level: see ref.~\cite{\Nonren}. 

It is common lore that $U(N)$ Yang-Mills theories on noncommutative 
Minowski are one-loop renormalizable. Indeed, if one assumes 
that gauge, better, BRS invariance is preserved at the one-loop level, 
it is difficult to think otherwise. 
However, statements about the BRS invariance of a field theory are 
rigorous only if they are based either on explicit computations or on the 
Quantum Action Principle~\cite{\QAP} plus BRS cohomology techniques  
\cite{\Algren}. Since we lack a Quantum Action Principle for noncommutative
field theories, we should better carry out explicit computations, lest our 
statements will  be erected on shaky ground. Even if we
had a Quantum Action Principle at our
 disposal, it would always be advisable to
check general results by performing explicit computations up to a few loops.  

In this paper we shall compute the complete UV divergent contribution to the 
1PI functional  of 4-dimensional noncommutative $U(N)$ Yang-Mills field theory 
for an arbitrary Lorentz gauge-fixing condition. We shall use dimensional 
regularization to carry out the computations. We shall thus  show by explicit 
computation that this 1PI functional is the sum of two integrated  
polynomials (with respect to the Moyal product)  of the  field and its 
derivatives. The first term is the noncommutative Yang-Mills action. This 
term is nontrivial in the cohomology of the noncommutative Slavnov-Taylor 
operator and gives rise to the renormalization of the coupling constant.  The
second term is exact in the cohomology of the noncommutative Slavnov-Taylor 
operator and gives rise to the wave function  and gauge-fixing
parameter renormalizations. This result constitutes a highly nontrivial check 
of the one-loop BRS invariance of the the theory; the high nontriviality of 
the check stemming from the fact that the UV divergence of each planar diagram
contributing to the 4-point  function of the gauge field has a structure
which differs very much from that of the 4-point tree-level contribution. 
The BRS invariance  of the MS (minimal subtraction) UV divergent 
part of the one-loop  1PI functional leads, as we shall see, to a 
renormalized BRS invariant 1PI functional up to order $\hbar$.

The layout of this paper is as follows. In Section 2, we set the notation   
and display the Feynman rules for noncommutative $U(N)$ 
Yang-Mills field theory in an arbitrary Lorentz gauge. In Section 3, 
we give the UV divergent divergent contribution to the one-loop 
1PI Green functions in the MS scheme of dimensional regularization and, 
from these data we obtain the MS UV divergent part of 1PI functional 
written in an explicitly BRS invariant form. Section 4 is devoted to 
comments and conclusions. In the Appendices we give for the record the 
UV divergent contribution to each one-loop 1PI diagram in the MS scheme.

\section{2.- Notation and Feynman rules}

The classical $U(N)$ field theory on noncommutative Minkowski space-time 
is given by the Yang-Mills functional
$$ 
 S_{YM}=-{1\over4 g^2\, \TR}\int\,\Trace( F_{\mu\nu}\star F^{\mu\nu}) (x),\eqno\numeq
$$\namelasteq\YMaction
where $F_{\mu\nu}(x)$ is given by
$$
F_{\mu\nu}(x)=\partial_{\mu}A_{\mu}-\partial_{\nu}A_{\mu}-i
\{A_{\mu},A_{\nu}\}(x),
$$ 
with $\{A_{\mu},A_{\nu}\}(x)=
(A_{\mu}\star A_{\nu})(x)-(A_{\nu}\star A_{\mu})(x)$.

The gauge field, $A_{\mu}$,  is an $N\times N$ hermitian matrix, 
$(A_\mu)^i_{\,j}=\sum_{a=0}^{N^2\!-1}A_\mu^a\,(T_a)^i_{\,j}$. We shall take  
the hermitian $U(N)$ generators, $T^a$,  normalized so that 
$\Trace T_a T_b = \TR \delta_{ab}$, if $a, b \ge 1$,
$T_0=t_{\rm R} \, 1_{N\times N}$ with $(t_{\rm R})^2\,N=\TR$.
$A_\mu^a$ is a real vector function on ${\rm {I\!R}}^4$. The symbol $*$ denotes
the Moyal, or star,  product defined as follows
$$
(f\star g)(x)=\int\int {d^4 q\over (2\pi)^4} {d^4 p\over (2\pi)^4}\;
e^{i (q+p)}\,  e^{i\,{1\over 2}\theta^{\mu\nu}q_{\mu}p_{\nu}}\;f(q)g(p).
$$
Here $f(q)$ and $q(p)$ are, respectively, the Fourier transforms of 
$f(x)$ and $g(x)$, the latter being two rapidly decreasing functions at infinity. $\theta^{\mu\nu}$ will be taken either magnetic or light-like, thus  
unitarity holds~\cite{\Unitarity}. 
The reader is referred to ref.~\cite{\Books} for introductions to  
Noncommutative geometry. We shall denote ${1 \over 2}\theta^{\mu\nu}\,p_{\mu}
q_{\nu}$ by $\omega(p,q)$.

To quantize the classical noncommutative $U(N)$ Yang-Mills theory we shall use
the path integral defined by means of the BRS quantization method. We 
introduce then the noncommutative ghost field, $c$, the noncommutative 
antighost field, $\bar{c}$, and the auxiliary field $B$. The fields $c$, 
$\bar{c}$ and $B$ are functions on ${\rm {I\!R}}^4$ with values on the 
Lie algebra of $U(N)$. The BRS transformations read thus
$$
sA_{\mu}(x)=D_{\mu}c(x)=\partial_{\mu}c(x)-i\{A_{\mu},c\}(x),\; 
s{\bar c}(x)=b,\; sb(x)=0,\; sc(x)=i(c\star c)(x). 
$$

The renormalization of the composite transformations 
$sA_{\mu}(x)$ and $sc(x)$ demands that the external fields
$J_{\mu}(x)$ and $H(x)$, which couple to them, be ushered in. 
The BRS invariant   $4$-dimensional classical action for an arbitrary 
Lorentz gauge-fixing condition reads
$$
S_{\rm cl} = S_{\rm YM} + S_{\rm gf} + S_{\rm ext},      \eqno\numeq
$$\namelasteq\BRSaction
where $S_{YM}$ has been given in eq.~\YMaction\ and
$$
\eqalignno{
S_{\rm gf} &= {1\over\TR}\,\int d^4x \;\Trace\, 
     s[{\bar c}\star({\lambda\over 2} B+
\partial_\mu A^\mu)](x),\cr
S_{\rm ext} &= \int d^4x \;\Trace\, 
\Bigl(J^\mu\star sA_\mu + H\star sc\Bigr)(x),       \cr
}
$$
where $\lambda$ is the gauge-fixing parameter.
 
Taking into account that
$$
\sum_{a=0}^{N^2-1}\,(T^a)^{j_1}_{\,i_1}\,(T^a)^{j_2}_{\,i_2}\, =
\,
 \TR\,\delta^{j_1}_{i_2}\,\delta^{j_2}_{i_1},
$$
one readily shows that the gauge field propagator reads
$$
\int {d^4 p\over (2\pi)^4} e^{-i p(x_1-x_2)}\; {(-i)\over p^2}
\,\TR\,g^2\,\delta_{i_1}^{j_2}\,\delta_{i_2}^{j_1}\;
\left[ g^{\mu_1\mu_2} + (\lambda'-1){p^{\mu_1} p^{\mu_2}\over p^2}\right]
\,,
$$
where $\lambda'\equiv\lambda/g^2$. The remaining propagators and tree-level
vertices are obtained from eq. \BRSaction\ by using standard path integral
techniques. We have gathered the Feynman rules in Figures 1 and 2. Note
that the Feynman rules are rendered in `t Hooft's double index 
notation~\cite{\Hooft}. This notation is very appropriate to work out 
the colour contribution to a given $U(N)$ Yang-Mills diagram.

\topinsert

{\settabs 4\columns \def\graphwidth{1.26in} 
\eightpoint

\+\hfil$\vcenter{\epsfxsize=\graphwidth\epsffile{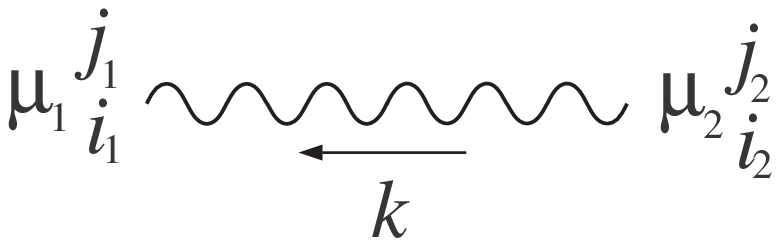}}$\hfil
 & $\vcenter{
     \hbox{$  G_{AA}^{(0)\;\mu1\mu2}{}^{j_1 j_2}_{i_1 i_2} (k_)=
  \;    \TR g^2\;\delta_{i_1}^{j_2} \delta_{i_2}^{j_1}\,
      {(-i)\over k^2}      
       \left[g^{\mu_1\mu_2}+(\lambda'-1){k^{\mu_1}k^{\mu_2}\over k^2}\right]
           $}
     }$\cr
\bigskip
\medskip

\+\hfil$\vcenter{\epsfxsize=\graphwidth\epsffile{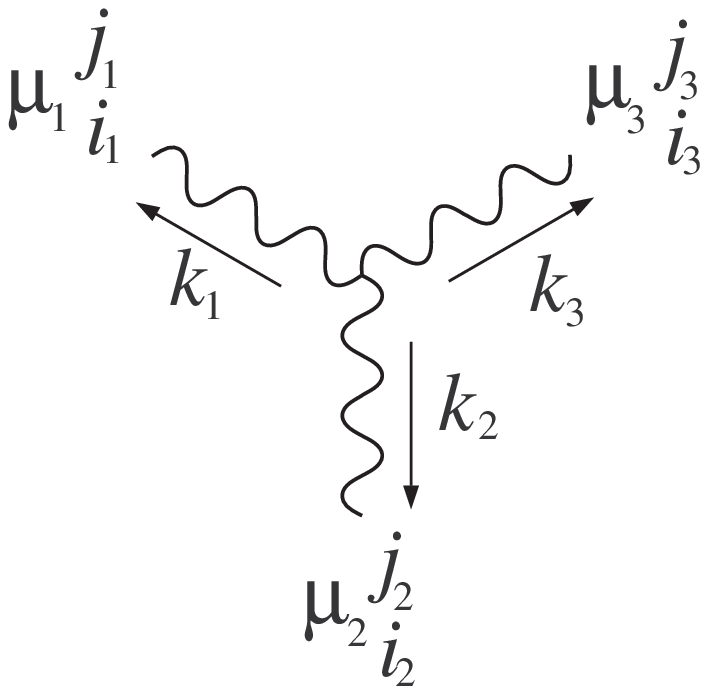}}$\hfil
 & $\vcenter{
      \hbox{$i\,   
            S_{AAA}^{\;\mu_1\mu_2\mu_3}{}^{ j_1 j_2 j_3}_{i_1 i_2 i_3}
            (k_1,k_2,k_3\!=\!-\!k_1\!-\!k_2)=
            $}
      \vskip 4pt
      \hbox{$\qquad 
              {i\over g^2 \TR} \,
              \left[ \delta_{i_1}^{j_3}\delta_{i_2}^{j_1}\delta_{i_3}^{j_2} 
                     \;e^{i\omega(k_1,k_2)} -
                     \delta_{i_1}^{j_2}\delta_{i_2}^{j_3}\delta_{i_3}^{j_1} 
                     \;e^{-i\omega(k_1,k_2)}
              \right]
            $}
      \vskip 2pt
      \hbox{$\qquad
              \left[g^{\mu_1\mu_2}(k_1-k_2)^{\mu_3} +
                    g^{\mu_2\mu_3}(k_2-k_3)^{\mu_1} +
                    g^{\mu_3\mu_1}(k_3-k_1)^{\mu_2}
              \right]
            $}
  }$\cr
\bigskip
\medskip
\+\hfil$\vcenter{\epsfxsize=\graphwidth\epsffile{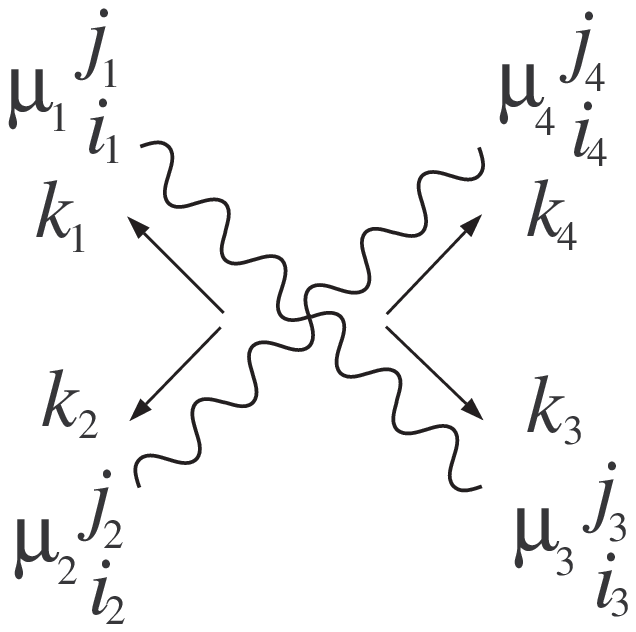}}$\hfil
 & $\vcenter{
      \hbox{$i\,  
            S_{AAAA}^{\;\mu_1\mu_2\mu_3\mu_4}
            {}^{ j_1 j_2 j_3 j_4}_{i_1 i_2 i_3 i_4}
            (k_1,k_2,k_3,k_4\!=\!-\!k_1\!-\!k_2\!-\!k_3)=
            $}
      \vskip 4pt
      \hbox{$\qquad 
              {i\over g^2 \TR}\; 
                     \delta_{i_1}^{j_4}\delta_{i_2}^{j_1}
                     \delta_{i_3}^{j_2}\delta_{i_4}^{j_3} 
                     \;e^{i[\omega(k_1,k_2)+\omega(k3,k4)]}
            $}
      \vskip 2pt
      \hbox{$\qquad
              \left(2 g^{\mu_1\mu_3}g^{\mu_2\mu_4}-
                      g^{\mu_1\mu_4}g^{\mu_2\mu_3}-
                      g^{\mu_1\mu_2}g^{\mu_3\mu_4}
              \right)
            $}
      \vskip 2pt
      \hbox{$\qquad + (1\,2\,4\,3)+(1\,3\,2\,4)+(1\,3\,4\,2)+
                      (1\,4\,2\,3)+(1\,4\,3\,2)
           $}
  }$\cr

\bigskip
\medskip

\+\hfil$\vcenter{\epsfxsize=\graphwidth\epsffile{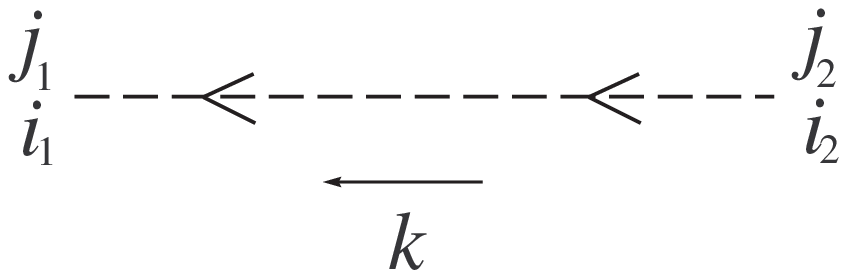}}$\hfil
  & $\vcenter{
        \hbox{$  G_{c \bar c}^{(0)}{}_{i_1 i_2}^{j_1 j_2}(k)=
    \;    \TR \;\delta_{i_1}^{j_2} \delta_{i_2}^{j_1}\;{i\over k^2} 
              $}
   }$\cr
\bigskip
\medskip

\+\hfil$\vcenter{\epsfxsize=\graphwidth\epsffile{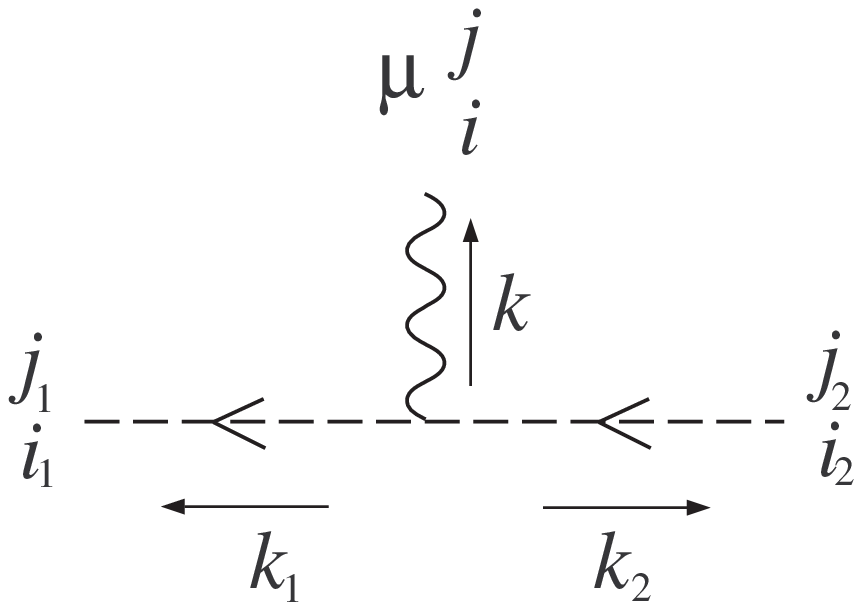}}$\hfil
  & $\vcenter{
        \hbox{$ i
                S_{c \bar c A}{}_{i_2 i_1 \,i}^{j_2 j_1 \,j\mu}
                     (k_2,k_1,k)=
              $}
	\vskip 4pt
        \hbox{$\qquad 
               {i k_1^\mu\over\TR}\;
               \left[e^{-i\omega(k_1,k_2)}\,
                        \delta_{i_1}^{j_2}\delta_{i_2}^j\delta_i^{j_1} -
                      e^{ i\omega(k_1,k_2)}\,
                        \delta_{i_1}^{j}\delta_{i_2}^{j_1}\delta_i^{j_2}
               \right]
              $}
    }$\cr
}
\vskip 12pt
\narrower\noindent {\bf Figure 1:}
{\eightpoint Feynman rules for noncommutative $U(N)$ Yang-Mills theory:
Propagators and vertices with no external fields. $(i\,j\,k\,l)$ denotes a 
permutation of $(1\,2\,3\,4)$.}

\vskip 0.1cm
\endinsert

\topinsert

{\settabs 4\columns \def\graphwidth{1.26in} 
\eightpoint

\+\hfil$\vcenter{\epsfxsize=\graphwidth\epsffile{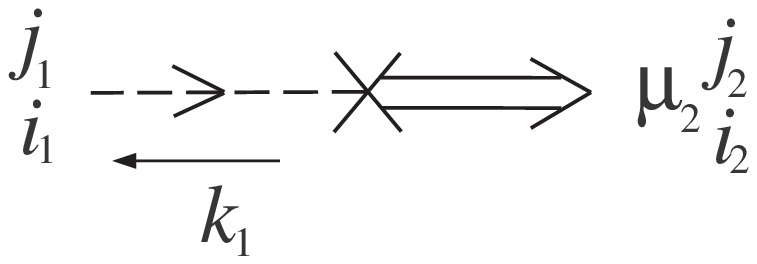}}$\hfil
 & $\vcenter{
     \hbox{$  S_{c;sA}^{(0)}{\,}^{j_1\, j_2}_{i_1; i_2\,\mu_2} (k_1)=
  \;    i\,{k_1}_{\mu_2}\;\delta_{i_1}^{j_2} \delta_{i_2}^{j_1}
           $}
     }$\cr
\bigskip
\medskip

\+\hfil$\vcenter{\epsfxsize=\graphwidth\epsffile{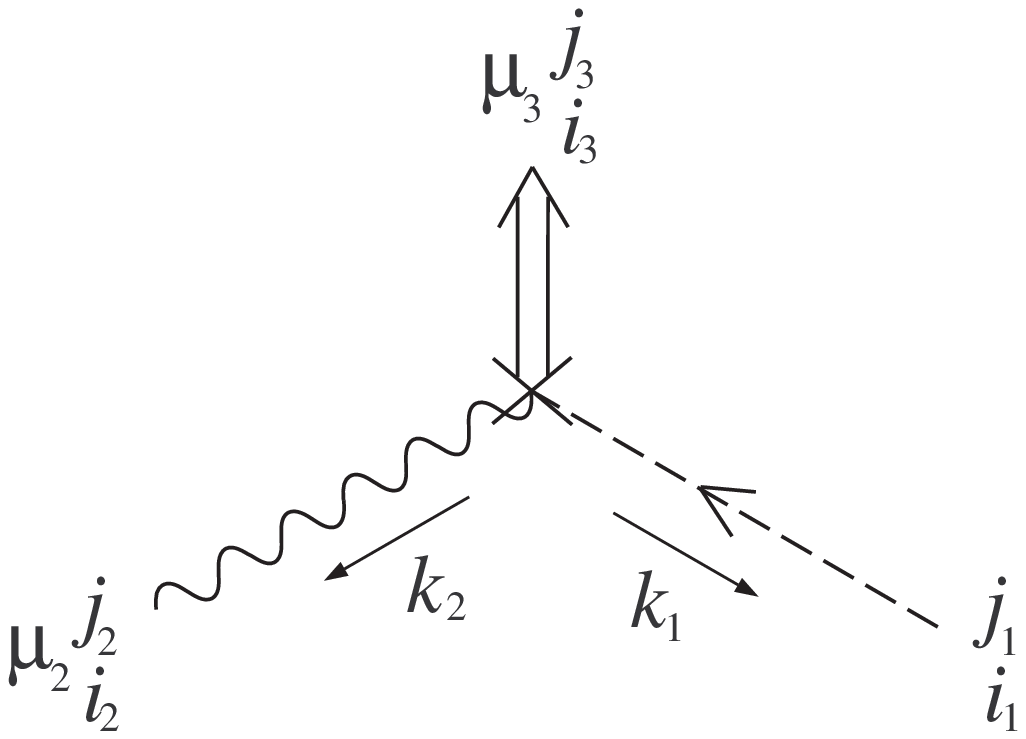}}$\hfil
 & $\vcenter{
      \hbox{$  S_{cA;sA}^{(0)}{\,}^{j_1 j_2}_{i_1i_2\mu_2;}
                                 {}^{j_3}_{i_3\mu_3}
            (k_1,k_2,k_3)=
            $}
      \vskip 4pt
      \hbox{$\qquad 
              {-i} \,
              \left[ \delta_{i_1}^{j_2}\delta_{i_2}^{j_3}\delta_{i_3}^{j_1} 
                     \;e^{-i\omega(k_1,k_2)} -
                     \delta_{i_1}^{j_3}\delta_{i_2}^{j_1}\delta_{i_3}^{j_2} 
                     \;e^{i\omega(k_1,k_2)}
              \right]\;g_{\mu_2\mu_3}
            $}
   }$\cr
\bigskip
\medskip
\+\hfil$\vcenter{\epsfxsize=\graphwidth\epsffile{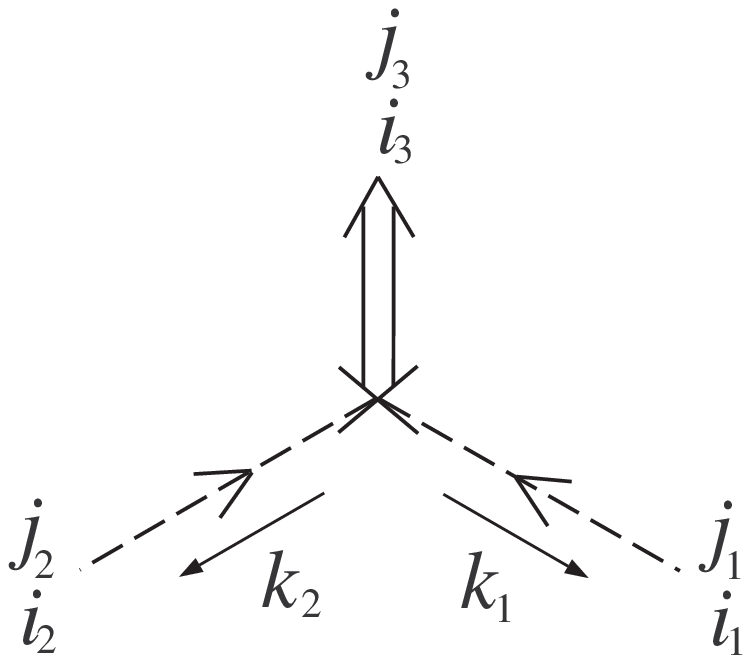}}$\hfil
 & $\vcenter{
      \hbox{$  S_{cc;sc}^{(0)}{\,}^{j_1 j_2\,j_3}_{i_1i_2;i_3}
            (k_1,k_2,k_3)=
            $}
      \vskip 4pt
      \hbox{$\qquad
              {i} \,
              \left[ \delta_{i_1}^{j_2}\delta_{i_2}^{j_3}\delta_{i_3}^{j_1} 
                     \;e^{-i\omega(k_1,k_2)} -
                     \delta_{i_1}^{j_3}\delta_{i_2}^{j_1}\delta_{i_3}^{j_2} 
                     \;e^{i\omega(k_1,k_2)}
              \right]
            $}
  }$\cr

}
\vskip 12pt
\narrower\noindent {\bf Figure 2:}
{\eightpoint Feynman rules for noncommutative $U(N)$ Yang-Mills theory:
Vertices with insertions of BRS variations.}

\vskip 0.1cm
\endinsert

 For the Noncommutative $U(N)$ gauge theory to be BRS invariant at the quantum
level the Slavnov-Taylor identity for the 1PI functional 
$\Gamma[A_\mu,B,c,{\bar c};J_\mu, H]$ must hold.
This identity reads
$$
{\cal S}(\Gamma)\equiv
\int d^4 x\, \;\Trace\,\Bigl[
{\delta\Gamma\over\delta J^\mu}
 {\delta\Gamma\over\delta A_\mu}+
 \,{\delta\Gamma\over\delta H}
 {\delta\Gamma\over\delta c}+
 \; B {\delta\Gamma\over\delta\bar c}\Bigr]=0. \eqno\numeq
$$\namelasteq\STidentity
For the one-loop, $\Gamma_1$, contribution to  
$\Gamma[A_\mu,B,c,{\bar c};J_\mu, H]$, the Slavnov-Taylor equation boils down
to
$$
{\cal B}\;\Gamma_{1}\,=\,0,\eqno\numeq
$$\namelasteq\oneST
where
$$
{\cal B}=
\int d^4 x\;{\rm Tr} \,\Bigl[
 \,{\delta S_{\rm cl}\over\delta J^\mu}
 {\delta \over\delta A_\mu}+
  {\delta S_{\rm cl}\over\delta A^\mu}
 {\delta\over\delta J_\mu}+
 \,{\delta S_{\rm cl}\over\delta H}
 {\delta \over\delta c}+{\delta S_{\rm cl}\over\delta c}
 {\delta\over\delta H}+
 \; B {\delta\over\delta\bar c}\Bigr]. \eqno\numeq
$$\namelasteq\FourBRSop
${\cal B}$ is the linearized noncommutative Slavnov-Taylor operator. 

Since the formal Feynman diagrams contributing to 
$\Gamma[A_\mu,B,c,{\bar c};J_\mu, H]$ present UV divergences, it is not 
straightforward that the renormalized, would it exist, 1PI functional 
defining the quantum theory satisfies the Slavnov-Taylor identity: anomalies
may occur. We shall see in this paper that at the one-loop level the theory 
at hand can be renormalized in such a way that eq.~\oneST\ holds for the 
MS renormalized action.

To regularize the Feynman integrals of our theory will shall use dimensional
regularization. To define $\theta^{\mu\nu}$ in dimensional regularization
we shall follow the philosophy in ref.~\cite{\BM} and say that 
$\theta^{\mu\nu}$ is an algebraic object which satisfies the following
equations
$$
\theta^{\mu\nu}=-\theta^{\nu\mu},\; \hat{\eta}_{\mu\rho}\theta^{\rho\nu}=0,
\;p_{\mu}\theta^{\mu\rho}\eta_{\rho\sigma}\theta^{\sigma\nu}p_{\nu}\geq 0, 
\forall p_{\mu}.
$$
Here $\eta_{\rho\sigma}$ and $\hat{\eta}_{\mu\rho}$ are, respectively, the
``D-dimensional'' and ``D-4-dimensional'' metrics as defined in 
ref.~\cite{\BM}. It is not difficult to convince oneself that, with the 
previous definition of $\theta_{\mu\nu}$, the one-loop Feynman integrals 
do have a mathematically well-defined expressions and that the techniques used 
in ref.~\cite{\BM} to prove the Quantum Action Principle for dimensionally 
regularized ordinary field theories can also be employed here to conclude
that at the one-loop level our dimensionally regularized noncommutative 
$U(N)$ is BRS invariant. The so regularized Feynman diagrams are meromorphic 
functions of $D$, with simple poles at $D=4$, if they are planar, and no poles,
if they are nonplanar and $P_{\mu}\theta^{\mu\rho}\eta_{\rho\sigma}\theta^{\sigma\nu}P_{\nu}> 0$ for any linear combination, $P$, of the external momenta.

\section{3.- The MS UV divergent part of the 1PI functional}

The fact that at the one-loop level only planar diagrams can be UV divergent 
and the fact that planar diagrams have the same  UV degree as their ordinary
field theory counterparts readily leads to the conclusion that the one-loop UV divergent 1PI functions are following: $\Gamma_{AA}$,  $\Gamma_{AAA}$, $\Gamma_{AAAA}$,
$\Gamma_{{\bar c} c}$,  $\Gamma_{{\bar c}A c}$, $\Gamma_{J c}$,
$\Gamma_{J A c}$ and $\Gamma_{Hcc}$, with obvious notation. 

We have computed the one-loop UV divergent 
contribution to all the divergent
1PI Green functions. Taking into account the results presented in the 
Appendices, one obtains the following values for these UV divergent part in
the MS scheme:
$$
\eqalignno{
 \Gamma_{\mu_1\mu_2}^{(AA),\,(pole)}{}{}_{i_1i_2}^{j_1j_2}(p)=&
   \cteI\;\Bigl({10\over 3} -(\lambda'-1) \Bigr)
   N\,\delta_{i_1}^{j_2}\,\delta_{i_2}^{j_1}
       \Bigl(p^2\,g_{\mu_1\mu_2}-p_{\mu_1}p_{\mu_2} \Bigr)\,\cr
 \Gamma_{\mu_1\mu_2\mu_3}^{(AAA),\, (pole)}{}{}_{i_1i_2i_3}^{j_1j_2j_3}(p_1,p_2,p_3)=&
\cteI\;\Bigl({4\over 3} -{3\over2}(\lambda'-1) \Bigr)\;\cr
   &N\,\left[e^{-i\omega(p_1,p_2)}
            \delta_{i_1}^{j_2}\,\delta_{i_2}^{j_3}\delta_{i_3}^{j_1}-
            e^{i\omega(p_1,p_2)}
            \delta_{i_1}^{j_3}\,\delta_{i_2}^{j_1}\delta_{i_3}^{j_2}
      \right]\cr
&\Bigl((p_1-p_2)_{\mu_3}g_{\mu_1\mu_2}+(p_2-p_3)_{\mu_1}g_{\mu_2\mu_3}+
(p_3-p_1)_{\mu_2}g_{\mu_1\mu_3}\Bigr)\,\cr
 \Gamma_{\mu_1\mu_2\mu_3\mu_4}^{(AAAA),\,(pole)}{}
{}_{i_1i_2i_3i_4}^{j_1j_2j_3j_4}
   (p_1,p_2,p_3,p_4)=&
  \cteI\;\Bigl({ 2\over 3} + 2(\lambda'-1) \Bigr)\cr
 &N\,
  \delta_{i_1}^{j_4}\,\delta_{i_2}^{j_1}\delta_{i_3}^{j_2}\delta_{i_4}^{j_3}
  \, e^{i[\omega(p_1,p_2) + \omega(p_3,p_4)]} \cr
 &(2g_{\mu_1\mu_3}g_{\mu_2\mu_4}-g_{\mu_1\mu_4}g_{\mu_2\mu_3}-
  g_{\mu_1\mu_2}g_{\mu_3\mu_4})\;+\cr
 &(1 2 4 3) + (1 3 2 4) + (1 3 4 2) + (1 4 2 3) + (1 4 3 2)\cr
 \Gamma^{(c{\bar c}),\,(pole)}{}{}_{i_2i_1}^{j_2j_1}(p_1)=&
  -\cteI\;g^2\;\Bigl(1-{\lambda'-1\over2}\Bigr)\,N\,
  \delta_{i_1}^{j_2} \delta_{i_2}^{j_1}\; p_1^{\,2} \cr
 \Gamma^{(c {\bar c}A ),\,(pole)}{}{}_{i_2i_1}^{j_2j_1}\,{}_{i_3\,\mu}^{j_3}
  (p_2,p_1,p_3)=&\cteI\,g^2\;\bigl(1+(\lambda'-1)\bigr)\cr
  &N\,\left[e^{-i\omega(p_1,p_2)}
            \delta_{i_1}^{j_2}\,\delta_{i_2}^{j_3}\,\delta_{i_3}^{j_1}-
            e^{i\omega(p_1,p_2)}
            \delta_{i_1}^{j_3}\,\delta_{i_2}^{j_1}\,\delta_{i_3}^{j_2}
      \right]\;p_{1\,\mu} \cr
 \Gamma^{(cJ),\,(pole)}{}{}_{i_1}^{j_1}{}_{i_2\,\mu_2}^{j_2}(p_1)=
  &-i\Bigl({1\over(4\pi)^2\varepsilon}\Bigr)
   \,g^2\,\TR\;\Bigl(1 -{\lambda'-1\over2}\Bigr) 
  \delta_{i_1}^{j_2}\,\delta_{i_2}^{j_1}\;p_{1\,\mu_2} \cr
 \Gamma^{(cAJ),\,(pole)}{}{\,}_{i_1}^{j_1}{}_{i_2\mu_2}^{j_2}
{}{}_{i_3\,\mu_3}^{j_3}
  (p_1,p_2,p_3)=
&-i\cteI\,\,g^2\,\TR\;\bigl(1+(\lambda'-1)\bigr)\cr
 &N\,\left[e^{-i\omega(p_1,p_2)}
            \delta_{i_1}^{j_2}\,\delta_{i_2}^{j_3}\,\delta_{i_3}^{j_1}-
            e^{i\omega(p_1,p_2)}
            \delta_{i_1}^{j_3}\,\delta_{i_2}^{j_1}\,\delta_{i_3}^{j_2}
      \right]\;g_{\mu_2\mu_3} \cr
 \Gamma^{(ccH),\,(pole)}{}{}_{i_1\,i_2\,i_3}^{j_1\,j_2\,j_3}(p_1,p_2,p_3)=
&i\Bigl({1\over(4\pi)^2\varepsilon}\Bigr)\,g^2\,\TR\;
   \bigl(1+(\lambda'-1)\bigr)\cr
 &N\,\left[e^{-i\omega(p_1,p_2)}
            \delta_{i_1}^{j_2}\,\delta_{i_2}^{j_3}\,\delta_{i_3}^{j_1}-
            e^{i\omega(p_1,p_2)}
            \delta_{i_1}^{j_3}\,\delta_{i_2}^{j_1}\,\delta_{i_3}^{j_2}
      \right]\,,      &\numeq\cr
}\namelasteq\Pole
$$
where $D=4-2\varepsilon$, $( i j k l)$ stands for the corresponding 
permutation of the indices and the following convention has been taken,
in order to take properly account of signs:
$$
\Gamma^{(\Phi_1\Phi_2\ldots\Phi_n)}_{\,i_1\,i_2\ldots i_n}=
  {\delta\,\Gamma\over\delta\Phi_1^{i_1}\delta\Phi_2^{i_2}\ldots
                      \delta\Phi_n^{i_n}}\bigg|_0\,,  \qquad
\Gamma^{(\phi_1\phi_2\ldots\phi_n K_\phi)}_{\,i_1\,i_2\ldots i_n i}=
  {\delta\,s\phi_i\cdot\Gamma\over\delta\phi_1^{i_1}\delta\phi_2^{i_2}\ldots
                      \delta\phi_n^{i_n}}\bigg|_0\,,
$$
with $\Phi$ standing for any, internal or external, field and $K_\phi$ 
for the external field coupled to $s\phi$. The already known results for the 
$U(1)$ theory are retrieved by setting $N=T_{R}=1$ in the previous expressions.

Notice that, upon formal generalization of $S_{\rm cl}$ in eq.~\BRSaction\ 
to the ``$D$-dimensional'' space of dimensional regularization,  the momentum 
structure of the singular contributions displayed above is the same
as the corresponding term in $S_{\rm cl}$. So, one  
would expect that these 1PI contributions can be subtracted by MS (minimal 
subtraction) multiplicative renormalization of the fields and parameters 
in the BRS invariant action. And, indeed, this is so, if we perform the 
following infinite renormalizations  
$$
\eqalign{&g_0={\mu}^{2\varepsilon}\,Z_g\,g,\;\lambda_0=\,Z_{\lambda}\,\lambda,\;
A_{0\,\mu}=\,Z_A\,A_{\mu},\;B_0=\,Z_{B}\,B,\;\cr
&J_{0\,\mu}=\,Z_J\,J_{\mu},\;H_0=\,Z_{H}\,H,\;
c_0=\,Z_{c}\,c \quad{\rm and}\quad{\bar c}_0=Z_{\bar c}\,{\bar c},\cr
}
$$
where the subscript $0$ labels the bare quantities. Now,
$$
\eqalignno{
S_{YM}^0= S_{YM} - &Z_g^{(1)}(2\,S_{AA}+2\,S_{AAA}+2\,S_{AAAA}) + \cr
                   &Z_A^{(1)}(2\,S_{AA}+3\,S_{AAA}+4\,S_{AAAA}), \cr 
}
$$
so that
$$
\eqalignno{
-2\,Z_g^{(1)}+2\,Z_A^{(1)}&={1\over(4\pi)^2\varepsilon}\, 2N\,
    \left[{5\over3}-{1\over2}(\lambda'-1)\right]\,g^2\TR,\cr
-2\,Z_g^{(1)}+3\,Z_A^{(1)}&={1\over(4\pi)^2\varepsilon}\, 2N\,
    \left[{2\over3}-{3\over4}(\lambda'-1)\right]\,g^2\TR,\cr
-2\,Z_g^{(1)}+4\,Z_A^{(1)}&={1\over(4\pi)^2\varepsilon}\, 2N\,
    \left[-{1\over3}-(\lambda'-1)\right]\,g^2\TR .\cr
}
$$
Hence,
$$
\eqalignno{
&Z_g=1-{1\over(4\pi)^2\varepsilon}\,2N\,{11\over 6}\,g^2\TR,\;
Z_A=1-{1\over(4\pi)^2\varepsilon}\,2N\,[1+{1\over4}(\lambda'-1)]\,g^2\TR
.&\numeq\cr
}\namelasteq\Zge
$$
Analogously, 
$$
\eqalignno{
&Z_{\bar c}Z_c=1+{1\over(4\pi)^2\varepsilon}\,{3-\lambda'\over 2}
    \,g^2\,\TR\,N,\;
Z_{\bar c}Z_A Z_c=1-{1\over(4\pi)^2\varepsilon}\,\lambda'\,g^2\,\TR\,N,\cr
&Z_H Z_{c}^2=1-{1\over(4\pi)^2\varepsilon}\,\lambda'\,g^2\,\TR\,N\,,\;
Z_B=Z_A^{-1},\;Z_{\lambda}=Z_A^2\quad {\rm and}\quad Z_J=Z_{\bar c}.&\numeq\cr
}\namelasteq\Zeds
$$

That the $Z$s in eqs.~\Zge\ and \Zeds\ render UV finite the one-loop 1PI functional is a consequence of BRS invariance. Indeed, in view of
eq.~\Pole, it is not difficult to show that the singular contribution, 
$\Gamma^{({\rm pole})}$, to the dimensionally regularized 1PI functional can be
cast into the form
$$
\Gamma^{(pole)}= 
{a\over 4g^2 T_R}\int d^Dx\; {\rm Tr}\;( F_{\mu\nu}\star F^{\mu\nu}) (x) + 
{\cal B}_{D}\, X, \eqno\numeq
$$\namelasteq\BRSpole
where
$$
\eqalign{ 
&X=\int d^Dx\; {\rm Tr}\; \Bigl(a_1 
(J_{\mu}+\partial_{\mu}{\bar c})\star A_{\mu}-
a_2 H\star c\Bigr)(x),\cr
a=&{1\over(4\pi)^2\varepsilon}\,{22\over 3}\, N\,T_R\,g^2,\;
a_1=+{1\over(4\pi)^2\varepsilon}\,{3+\lambda'\over 2}\,N\,T_R\,g^2,\;
a_2=+{1\over(4\pi)^2\varepsilon}\,\lambda'\,N\,T_R\,g^2,\cr
}
$$
and ${\cal B}_{D}$ is the linearized Slavnov-Taylor operator acting upon the
space of formal $*$-polynomials  constructed with ``$D$-dimensional'' 
monomials of the fields and their derivatives. 
${\cal B}_{D}$ is defined as follows
$$
{\cal B}_{D}=
\int d^D x\; {\rm Tr}\;\Bigl[
 \,{\delta S_{\rm cl}\over\delta J^\mu}
 {\delta \over\delta A_\mu}+
  {\delta S_{\rm cl}\over\delta A^\mu}
 {\delta\over\delta J_\mu}+
 \,{\delta S_{\rm cl}\over\delta H}
 {\delta \over\delta c}+{\delta S_{\rm cl}\over\delta c}
 {\delta\over\delta H}+
 \; B {\delta\over\delta\bar c}\Bigr]. 
$$
The operator ${\cal B}_{D}$ is the ``$D$-dimensional'' counterpart of the
operator $\cal B$ defined in eq.~\FourBRSop.

Eq.~\BRSpole\ gives, in an explicitly BRS invariant form, the UV divergent 
contribution to the 1PI functional in the MS scheme of  Dimensional 
Regularization. Note that $\Gamma^{({\rm pole})}$ is the sum of two terms: 
the first term, the Yang-Mills term, is ${\cal B}_D$-closed and the 
second term is ${\cal B}_D$-exact (recall that ${\cal B}_D^2=0$). The 
analogy with ordinary $SU(N)$ Yang-Mills theory is apparent. And, indeed, as  
as in ordinary four-dimensional Yang-Mills theory we have
$$ 
\eqalign{ &Z_g=1-{a\over 2},\; Z_A=1+a_1,\; 
Z_{\bar c}Z_c=1-a_1+a_2,\;
Z_{\bar c}Z_A Z_c=1+a_2,\cr 
&Z_H Z_{c}^2=1+a_2,\;Z_B=Z_A^{-1},\;Z_{\lambda}=
Z_A^2\quad {\rm and}\quad Z_J=Z_{\bar c}.\cr
}
$$
Eqs.~\Zge\ and \Zeds\ are thus retrieved. Let us remark that the values 
we have obtained for the $Z$s agree with the corresponding values of the 
$Z$s of $SU(N)$ Yang-Mills theory on ordinary Minkowski space-time.

We shall define as usual the renormalized one-loop 1PI functional,  
$\Gamma_{\ren}^{(1),\,{\rm MS}}$, in the MS scheme:
$$
\Gamma_{\ren}^{(1),\,{\rm MS}}=
{\cal LIM}_{\varepsilon\rightarrow 0}
\Bigl[\Gamma^{(1)}_{{\rm DReg}}-\Gamma^{({\rm pole})}\Bigr].
$$
Here $\Gamma^{(1)}_{{\rm DReg}}$ denotes the dimensionally regularized
1PI functional at order $\hbar$ and  the functional $\Gamma^{({\rm pole})}$ 
is given in eq.~\BRSpole. The limit $\varepsilon\rightarrow 0$ is taken after 
performing the subtraction of the pole  and replacing 
every ``$D$-dimensional'' algebraic object with its
4-dimensional counterpart~\cite{\BM}; this is 
 why we have denoted it by ${\cal LIM}$. We shall not discuss in this paper
how to  make sense out of the noncommutative IR divergences~\cite{\MRS} that  
occur in $\Gamma_{\ren}^{(1),\,{\rm MS}}$. 

Since  $\Gamma^{(1)}_{{\rm DReg}}$ is 
BRS invariant, i.e., it satisfies the Slavnov-Taylor identity 
at order $\hbar$
$$
{\cal B}_{D}\Gamma^{(1)}_{{\rm DReg}}=0,
$$
the MS renormalized 1PI functional $\Gamma_{\ren}^{(1),\,{\rm MS}}$ is
also BRS invariant:
$$
{\cal B}\,\Gamma_{\ren}^{(1),\,{\rm MS}}=0.
$$
The operator ${\cal B}$ has been defined in eq.~\FourBRSop.
We thus conclude that the Slavnov-Taylor identity (eq.~\STidentity)  holds 
for the renormalized theory at order $\hbar$. 

By using standard textbook techniques, one can work out the
renormalization group equation -expressing the invariance of the observables 
under changes of the renormalization scale $\mu$- for $\Gamma_{\ren}^{\rm MS}$:
$$
\Bigl[\mu{\partial \over \partial\mu}+\beta{\partial \over \partial g^2}
-\delta_{\lambda}{\partial \over \partial \lambda}-\sum_{\phi}\gamma_{\phi}\,
\int d^4 x\,\phi(x){\delta \over \delta\phi(x)}\Bigr]\,
\Gamma^{\rm MS}_{\ren}[\phi;g,\theta,\lambda]=0.
$$
The fields are denoted  by $\phi$.    
It should be noticed that $\theta_{\mu\nu}$ is a dimensionful parameter 
which is not renormalized in the MS renormalization scheme.
 
The one-loop beta function of the theory is easily computed to be
$$
\beta(g^2)\equiv \mu{d g^2\over d\mu}= -{1\over 8\pi^2}\,{22\over 3}
\,N\,T_R\,g^4.
$$
The other renormalization group coefficients read at the one-loop level:
$$
\eqalign{
&\gamma_{A}=+{1\over 8\pi^2}\big({3+\lambda'\over 2}\bigr)\,N\,T_R\,g^2,\quad
\gamma_{c}=+{1\over 8\pi^2}\,\lambda'\,N\,T_R\,g^2,\cr
&\gamma_{J}=\gamma_{\bar c}=\gamma_{B}=-\,\gamma_{A},\quad 
\delta_{\lambda}=-\,2\,\gamma_{A}\,\lambda,\quad\gamma_{H}=-\,\gamma_{c}.\cr
}
$$

In the proof of the renormalizability of ordinary $SU(N)$ Yang-Mills theory,   
besides the Slavnov-Taylor equation, two equations play
an important role, namely, the gauge-fixing and ghost equation~\cite{\Algren}.
Do these equations also hold for noncommutative $U(N)$ Yang-Mills theory? 
It is not difficult to show that   $\Gamma^{({\rm pole})}$ in eq.~\BRSpole\ 
verifies both the gauge-fixing equation and the ghost equation:
$$
{\delta \Gamma^{({\rm pole})}\over\delta B}=0,\quad
{\delta \Gamma^{({\rm pole})}\over\delta \bar c}+
\partial_{\mu}{\delta \Gamma^{({\rm pole})}\over\delta J_{\mu}}=0.
$$ 
Hence, up to order $\hbar$, the MS renormalized 1PI functional does 
satisfy both the gauge-fixing equation and the ghost equation:
$$
\eqalign{T_R &{\delta \Gamma^{\rm MS}_{\ren}\over\delta B}=
\lambda B + \partial A\,+\,O(\hbar^2),\cr
&{\delta \Gamma^{\rm MS}_{\ren}\over\delta \bar c}+
\partial_{\mu}{\delta \Gamma^{\rm MS}_{\ren}\over\delta J_{\mu}}=0
\,+\,O(\hbar^2).\cr
}
$$

\section{4.- Conclusions and comments}

In this paper we have computed, within the MS scheme of dimensional 
regularization, all the one-loop UV divergent contributions to the 
1PI functional for noncommutative $U(N)$ Yang-Mills theory in an arbitrary 
Lorentz gauge. We have shown that these contributions satisfy the 
Slavnov-Taylor equation, the gauge-fixing equation and the ghost equation and,
hence, that the MS renormalized 1PI functional 
satisfies those equations as well. That the one-loop UV divergent 
contributions to the 1PI functional satisfy
the Slavnov-Taylor equation has been shown by casting the pole part of the
1PI functional in an explicitly BRS invariant form. As with ordinary $SU(N)$ 
Yang-Mills theory, this explicitly BRS invariant form is the sum of a 
${\cal B}$-closed term -which is proportional to the Yang-Mills action- and
${\cal B}$-exact term; being ${\cal B}$ the linearized noncommutative BRS
operator. Our computations lead us to believe that the $*$-deformation of
the ordinary BRS cohomology techniques~\cite{\Algren} are to play an 
important role in the proof of the perturbative renormalizability of 
noncommutative gauge theories. 

That the 2- and 3-point functions of noncommutative $U(N)$ theory can be 
renormalized in a BRS invariant way does not come as a surprise once one 
shows, as we have done in this paper, that the sum of a given diagram 
with its permutations, the diagrams being planar, is proportional to the 
tree-level 1PI Green function. This 
generalizes the results in refs.~\cite{\Tony, \Filk}.
The reader is referred to the Appendices for further details; where he can
realize, in particular, that every contribution listed there grows with
$N$ as corresponds to the fact that they come only from planar diagrams.

That the one-loop 4-point function of the gauge field has a MS UV divergent 
part which does not spoil the Slavnov-Taylor equation is, though, highly 
nontrivial. It demands that very delicate cancellations occur upon adding 
all the UV divergent 4-point diagrams: unlike 2- and 3-point diagrams, 
the sum of 4-point diagrams of the same type is not proportional to
tree-level 4-point gauge vertex. That these delicate cancellations 
do happen is, of course, a consequence of the fact that BRS invariance 
holds indeed.   In this regard, we invite the reader to go  to 
Appendix C and eq.~\Pole\ and get acquainted with the momentum and 
colour structure of the UV divergent contributions reported there. Note that
every 4-point  UV divergent contribution has an overall factor equal to $N$,
for they come  only from planar diagrams.

As regards the actual value of the beta function, the anomalous dimensions 
of the fields and gauge-fixing parameter, we have found that they are
those of $SU(N)$'s. This result is, of course, almost 
trivial~\cite{\Filk, \MRS}, once it is shown that BRS holds. Indeed, 
all 1PI planar diagrams but the 4-point diagrams can be grouped in classes of
planar diagrams of the same type, the sum of the diagrams in each class
being proportional to the corresponding tree-level contributions. 
However, taking into account what it is at stake, computations which are 
both explicit and thorough are much welcomed.

Finally, the computations presented in this paper were finished more than 
a year ago. In the meantime two papers which overlap with it 
have appeared~\cite{\NCUN}. Our findings are in agreement with theirs, but  
ours are more general.

\bigskip

\section{Appendix A. Gauge field 2-point function}

The diagrams which are UV divergent in dimensional regularization are the 
planar diagrams in Figure 3. Note that the planar tadpole diagram is not
singular at $D=4$ in dimensional regularization.

\midinsert
{\settabs 2\columns \def\graphwidth{1.7in}   
\eightpoint
\+\hfill\epsfxsize=\graphwidth\epsffile{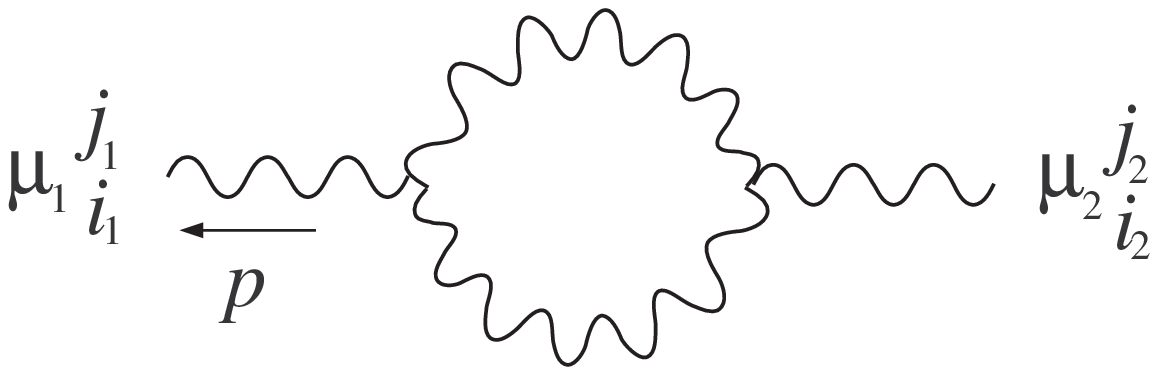}\hfill
	&\hfill\epsfxsize=\graphwidth\epsffile{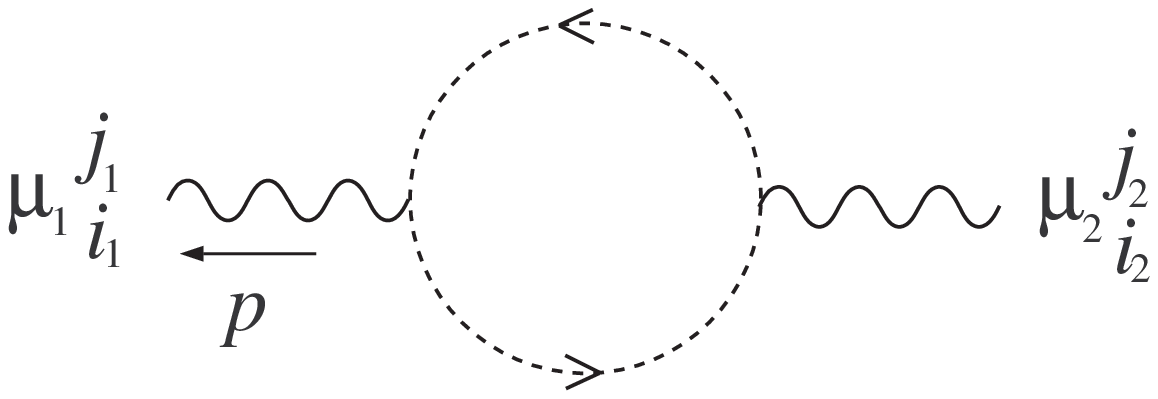}\hfill &\cr
\+\hfill$(i)$\hfill
	& \hfill $(ii)$\hfill &\cr
}

\vskip 12pt
\narrower\noindent {\bf Figure 3:}
{\eightrm 1PI UV divergent  2-point Feynman diagrams for the gauge field.}
\vskip 0.4cm
\endinsert
The minimal UV divergent part of each diagram reads
$$
\eqalignno{
(i)=&
   2i\cteI\, 	 N\,\delta_{i_1}^{j_2}\,\delta_{i_2}^{j_1}
\Bigg[\Bigl({19\over12} -{(\lambda'-1)\over2} \Bigr)
        p^2\,g_{\mu_1\mu_2} -
      \Bigl({11\over6} -{(\lambda'-1)\over2} \Bigr)
        p_{\mu_1}p_{\mu_2} 
\Bigg], \cr
(ii)=&
   2i\cteI\, 	 N\,\delta_{i_1}^{j_2}\,\delta_{i_2}^{j_1}
\Bigg[{1\over12}p^2\,g_{\mu_1\mu_2} +
      {1\over6}p_{\mu_1}p_{\mu_2} 
\Bigg]\,.\cr
}
$$

\bigskip

\section{Appendix B. Gauge field 3-point function}

The UV divergent part of the 3-point function of the gauge field is 
obtained from the planar diagrams in Figure 4.
\midinsert
{\settabs 3\columns \def\graphwidth{1.5in}   
 \def\graphwidthbis{1in}
\eightpoint
\+\hfill\epsfxsize=\graphwidthbis\epsffile{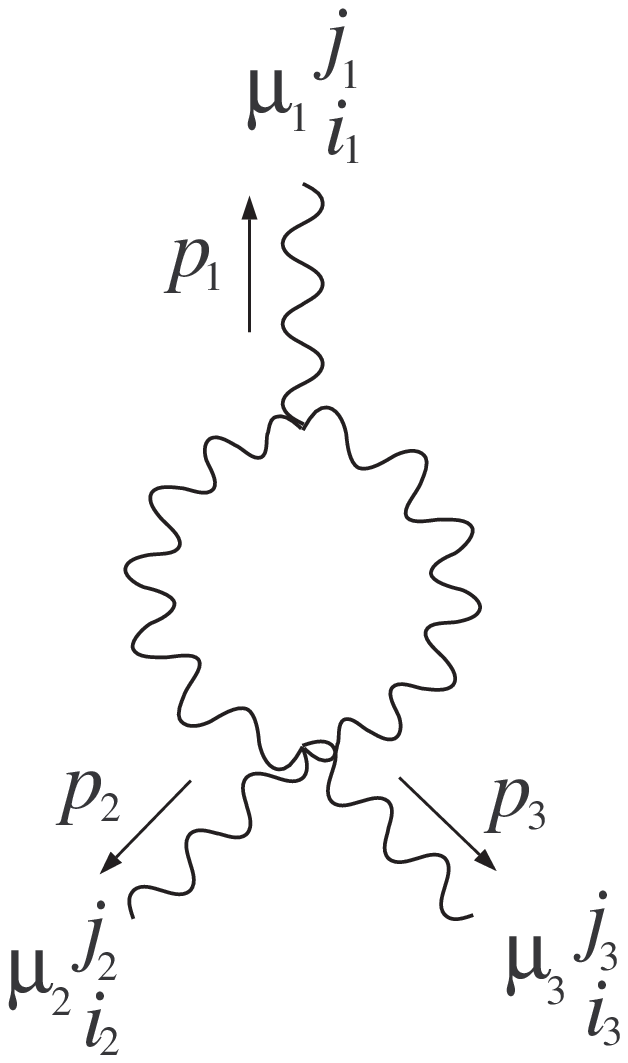}\hfill
	&\hfill\epsfxsize=\graphwidth\epsffile{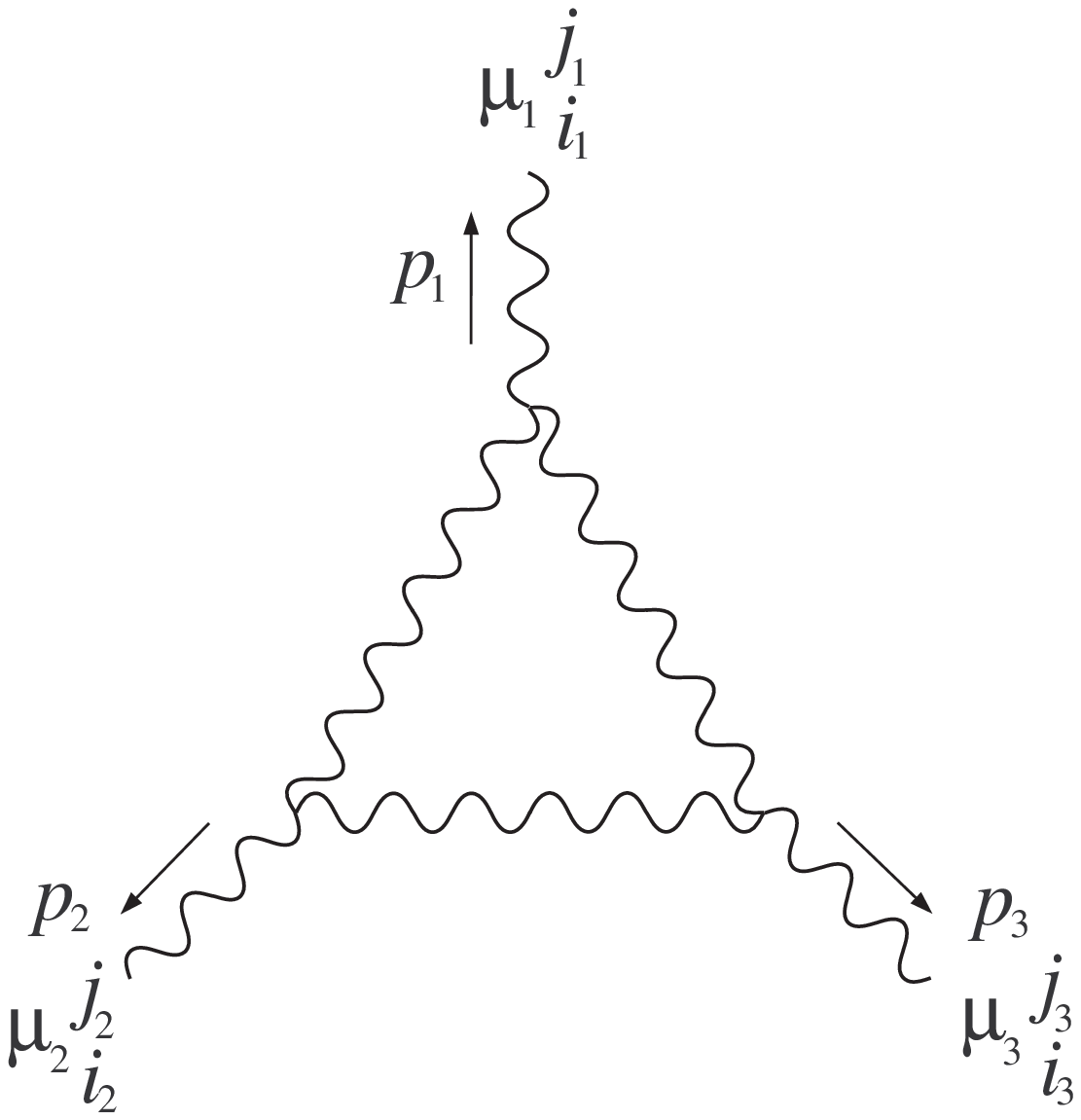}\hfill
	&\hfill\epsfxsize=\graphwidth\epsffile{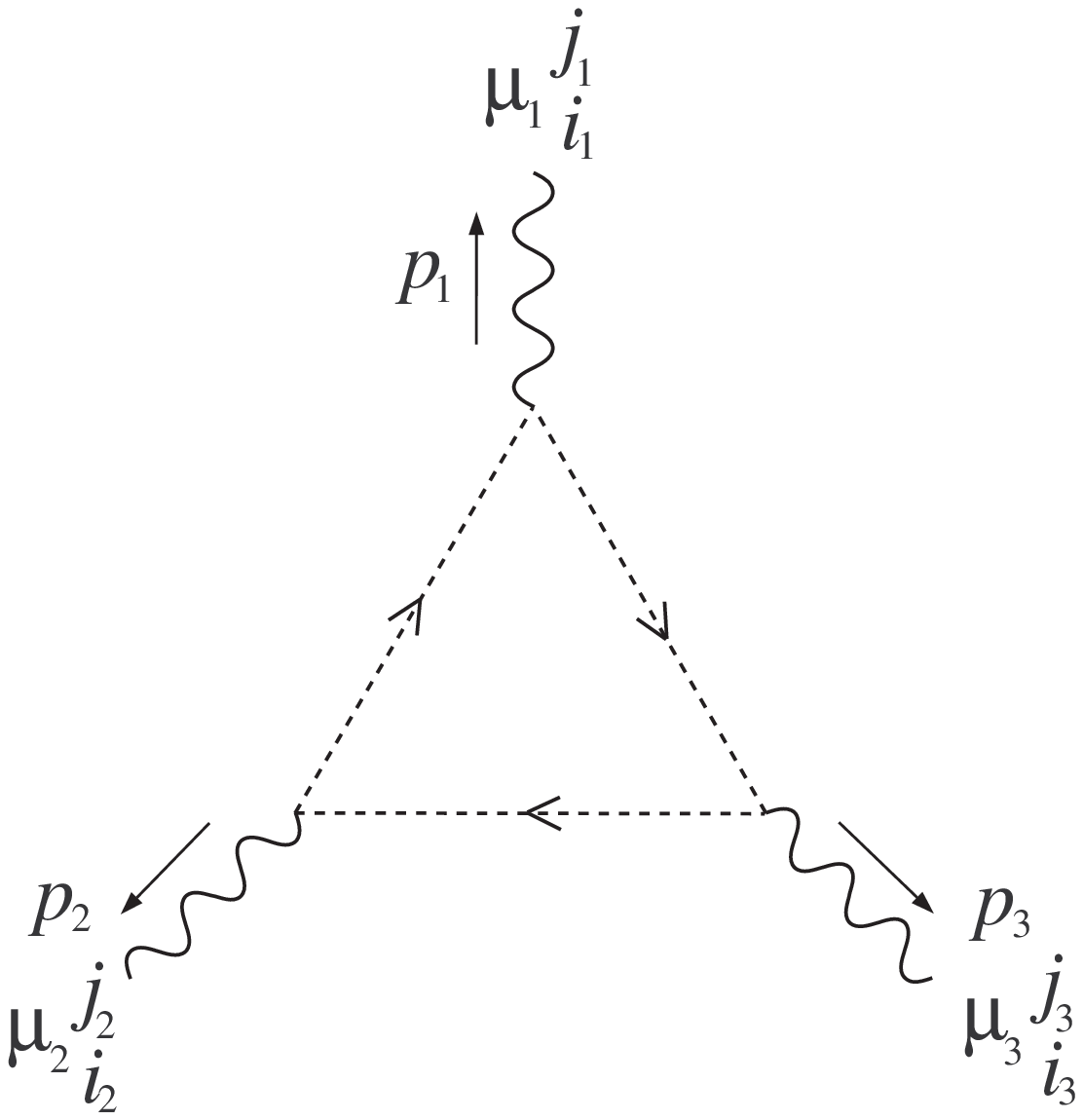}\hfill &\cr
\+\hfill$(i)_{123}$\hfill
	&\hfill $(ii)_{123}$\hfill
	& \hfill $(iii)_{123}$\hfill &\cr
}

\vskip 12pt
\narrower\noindent {\bf Figure 4:}
{\eightrm 1PI UV divergent 3-point Feynman diagrams for the gauge field.}
\vskip 0.4cm
\endinsert
The UV divergent part of these diagrams read
$$
\eqalignno{
(i)_{123}=&
i\cteI\;\Bigl({9\over2} +{3(\lambda'-1)\over4} \Bigr)\;
 N\,\left[e^{-i\omega(p_1,p_2)}
            \delta_{i_1}^{j_2}\,\delta_{i_2}^{j_3}\delta_{i_3}^{j_1}-
            e^{i\omega(p_1,p_2)}
            \delta_{i_1}^{j_3}\,\delta_{i_2}^{j_1}\delta_{i_3}^{j_2}
      \right]\,\times\cr
&\Bigl({p_1}_{\mu_3}g_{\mu_1\mu_2}-{p_1}_{\mu_2}g_{\mu_1\mu_3}
 \Bigr),\cr
(ii)_{123}=&
i\cteI\;\Bigl({13\over 4} +{9(\lambda'-1)\over4} \Bigr)\;
 N\,\left[e^{-i\omega(p_1,p_2)}
            \delta_{i_1}^{j_2}\,\delta_{i_2}^{j_3}\delta_{i_3}^{j_1}-
            e^{i\omega(p_1,p_2)}
            \delta_{i_1}^{j_3}\,\delta_{i_2}^{j_1}\delta_{i_3}^{j_2}
      \right]\,\times\cr
&\Bigl((p_2-p_1)_{\mu_3}g_{\mu_1\mu_2}-(p_1+2p_2)_{\mu_1}g_{\mu_2\mu_3}+
(p_2+2p_1)_{\mu_2}g_{\mu_1\mu_3}\Bigr),\cr
(iii)_{123}=&
i\cteI\;{1\over12}\;
   N\,\left[e^{-i\omega(p_1,p_2)}
            \delta_{i_1}^{j_2}\,\delta_{i_2}^{j_3}\delta_{i_3}^{j_1}-
            e^{i\omega(p_1,p_2)}
            \delta_{i_1}^{j_3}\,\delta_{i_2}^{j_1}\delta_{i_3}^{j_2}
      \right]\,\times\cr
&\Bigl((-p_1-2p_2)_{\mu_3}g_{\mu_1\mu_2}+(2p_1+p_2)_{\mu_1}g_{\mu_2\mu_3}+
(p_2-p_1)_{\mu_2}g_{\mu_1\mu_3}\Bigr)\,.\cr
}
$$

Summing over permutations one obtains
$$
\eqalignno{
(I)=&(i)_{123} + (i)_{231} + (i)_{312}=\cr
&i\cteI\;\Bigl({9\over2} +{3(\lambda'-1)\over4} \Bigr)\;
 N\,\left[e^{-i\omega(p_1,p_2)}
            \delta_{i_1}^{j_2}\,\delta_{i_2}^{j_3}\delta_{i_3}^{j_1}-
            e^{i\omega(p_1,p_2)}
            \delta_{i_1}^{j_3}\,\delta_{i_2}^{j_1}\delta_{i_3}^{j_2}
      \right]\,\times\cr
&\Bigl((p_1-p_2)_{\mu_3}g_{\mu_1\mu_2}-(2p_1+p_2)_{\mu_2}g_{\mu_1\mu_3}
  + (p_1+2p_2)_{\mu_1}g_{\mu_2\mu_3}
 \Bigr),\cr
(II)=&(ii)_{123}=\cr
&i\cteI\;\Bigl({13\over 4} +{9(\lambda'-1)\over4} \Bigr)\;
 N\,\left[e^{-i\omega(p_1,p_2)}
            \delta_{i_1}^{j_2}\,\delta_{i_2}^{j_3}\delta_{i_3}^{j_1}-
            e^{i\omega(p_1,p_2)}
            \delta_{i_1}^{j_3}\,\delta_{i_2}^{j_1}\delta_{i_3}^{j_2}
      \right]\,\times\cr
&\Bigl((p_2-p_1)_{\mu_3}g_{\mu_1\mu_2}-(p_1+2p_2)_{\mu_1}g_{\mu_2\mu_3}+
(p_2+2p_1)_{\mu_2}g_{\mu_1\mu_3}\Bigr),\cr
(III)=&(iii)_{123}+(iii)_{213}=\cr
&i\cteI\;{1\over12}\;
   N\,\left[e^{-i\omega(p_1,p_2)}
            \delta_{i_1}^{j_2}\,\delta_{i_2}^{j_3}\delta_{i_3}^{j_1}-
            e^{i\omega(p_1,p_2)}
            \delta_{i_1}^{j_3}\,\delta_{i_2}^{j_1}\delta_{i_3}^{j_2}
      \right]\,\times\cr
&\Bigl((p_1-p_2)_{\mu_3}g_{\mu_1\mu_2}+(p_1+2p_2)_{\mu_1}g_{\mu_2\mu_3}+
(-2p_1-p_2)_{\mu_2}g_{\mu_1\mu_3}\Bigr).\cr
}
$$
The sum of $(I)$, $(II)$ and $(III)$ yields the UV divergent contribution
to the 1PI 3-point function of the gauge field. Note that (I), (II) and (III)
are proportional to the 3-point vertex of the gauge field.

\bigskip

\section{Appendix C. Gauge field 4-point function}

The UV divergent contribution to the 4-point 1PI function of the gauge field
is computed by summing over the UV divergent parts of the diagrams, and the 
appropriate permutations of these diagrams, in Figure 5.
\midinsert
{\settabs 4\columns \def\graphwidth{1.5in}   
 \def\graphwidthbis{1.1in}
\eightpoint
\+\hfill\epsfxsize=\graphwidthbis\epsffile{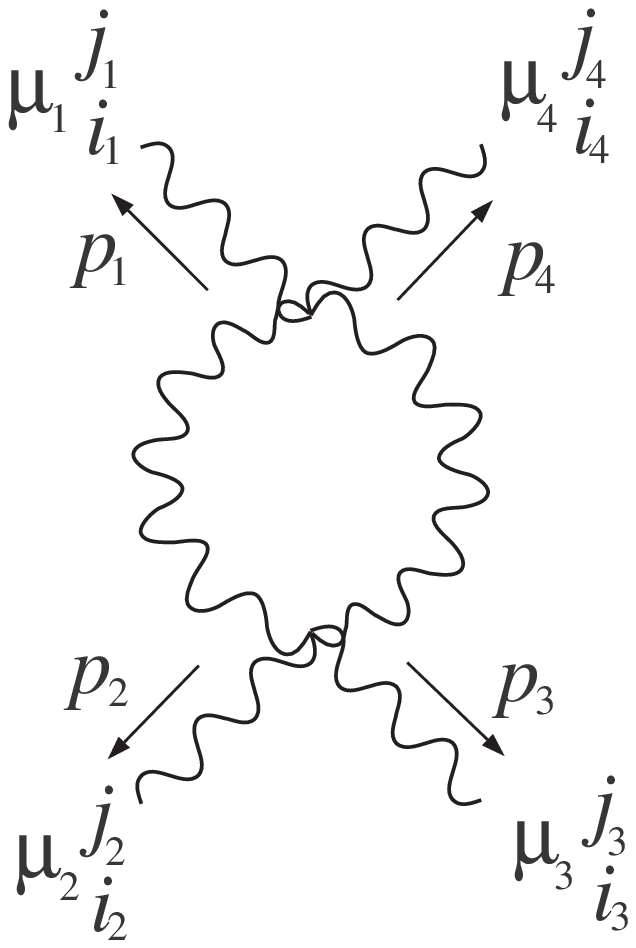}\hfill
	&\hfill\epsfxsize=\graphwidth\epsffile{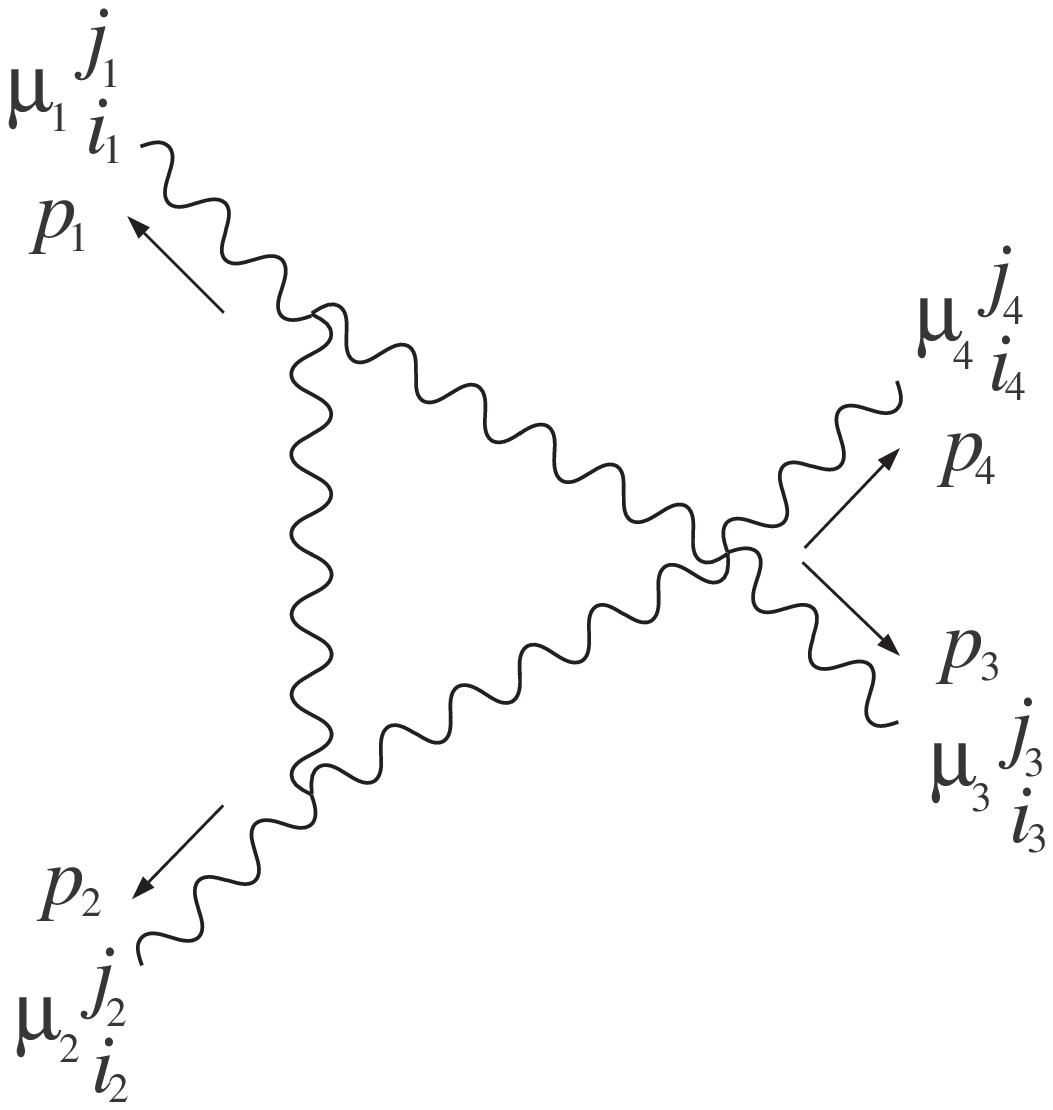}\hfill
	&\hfill\epsfxsize=\graphwidth\epsffile{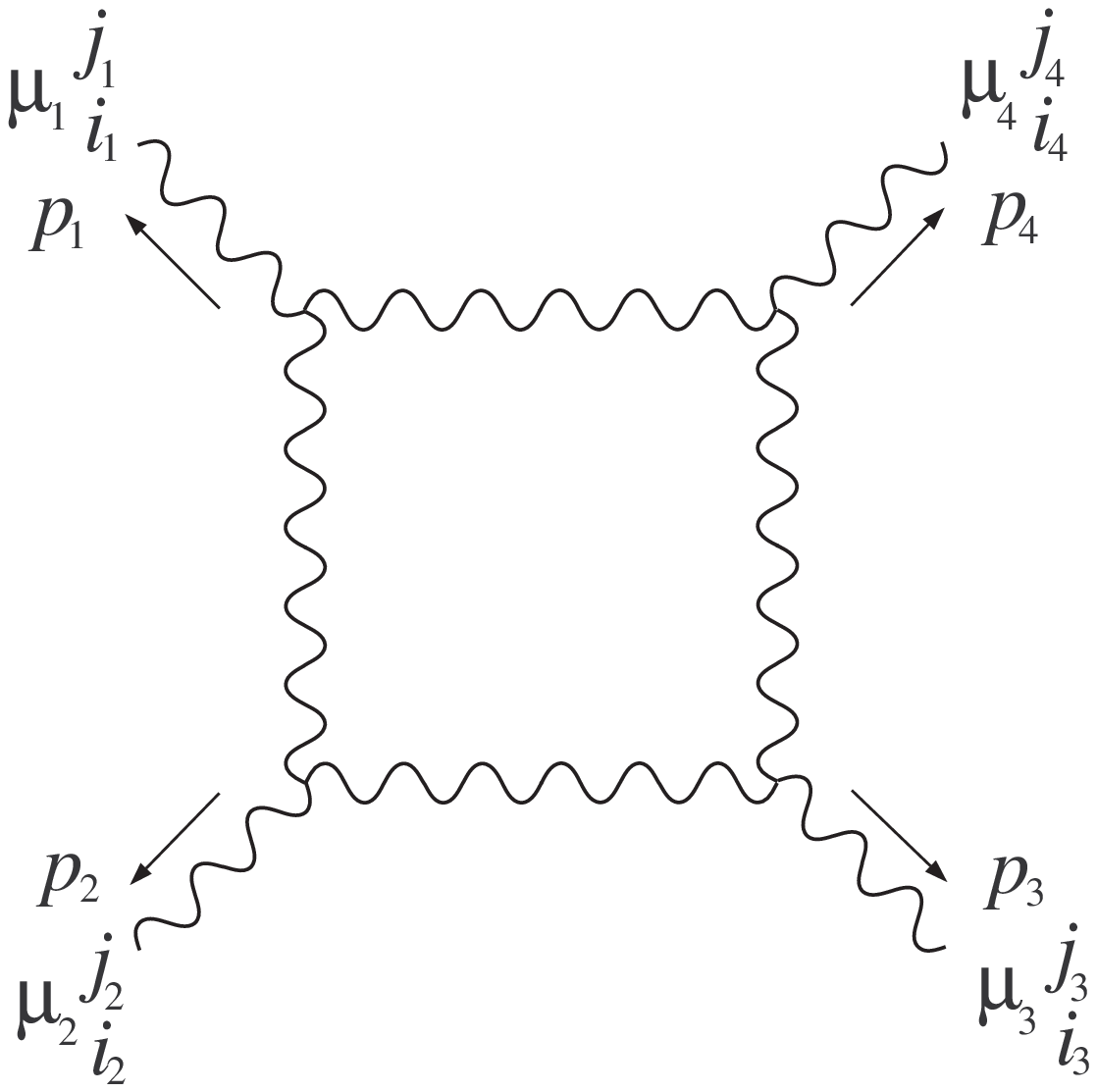}\hfill 
	&\hfill\epsfxsize=\graphwidth\epsffile{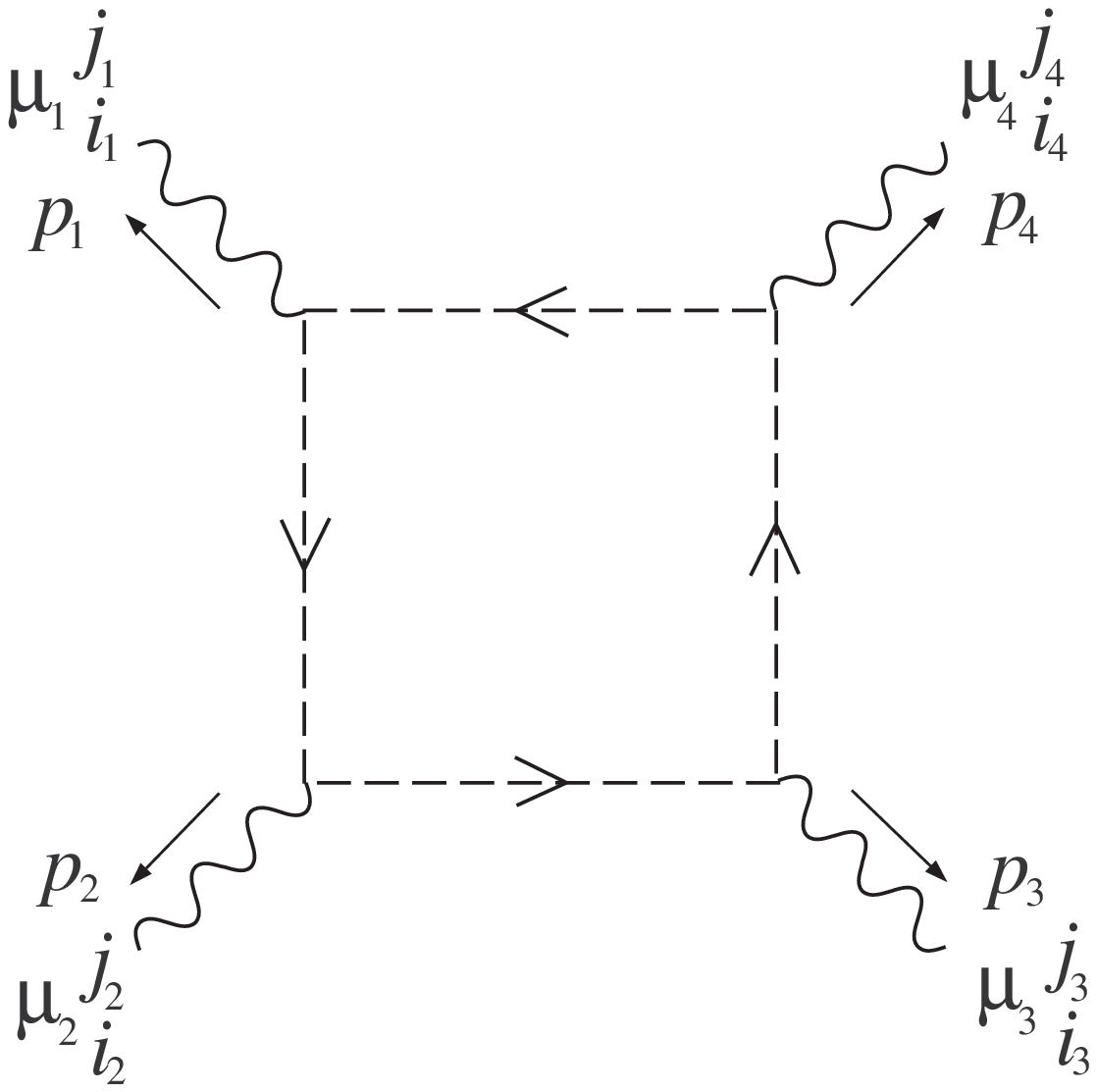}\hfill &\cr
\+\hfill$(i)_{1234}$\hfill
	&\hfill $(ii)_{1234}$\hfill
	&\hfill $(iii)_{1234}$\hfill
	& \hfill $(iv)_{1234}$\hfill &\cr
}

\vskip 12pt
\narrower\noindent {\bf Figure 5:}
{\eightrm 1PI UV divergent 4-point Feynman diagrams for the gauge field.}
\vskip 0.4cm
\endinsert
The UV divergent part of the diagrams in Figure 5 read
$$
\eqalignno{ 
(i)_{1234}=i\,&\cteI\,N\;\times\cr
\Bigg\{\!
  &\left(
   e^{i[-\omega(p_1\!,p_2\!) - \omega(p_1\!,p_3\!)+ \omega(p_2\!,p_3\!)]}
   \delta_{i_1}^{j_3}\delta_{i_2}^{j_4}\delta_{i_3}^{j_2}\delta_{i_4}^{j_1}
   \!+
   e^{i[\omega(p_1\!,p_2\!) + \omega(p_1\!,p_3\!)- \omega(p_2\!,p_3\!)]}
   \delta_{i_1}^{j_4}\delta_{i_2}^{j_3}\delta_{i_3}^{j_1}\delta_{i_4}^{j_2}
   \right)\cr 
    &\quad\bigg[\left(2+(\lambda'-1)+{13(\lambda'-1)^2\over24}\right) 
            g_{\mu_1\mu_4}\,g_{\mu_2\mu_3} +\cr
    &\qquad\left(5+{5(\lambda'-1)\over2}+{(\lambda'-1)^2\over24}\right) 
            g_{\mu_1\mu_3}\,g_{\mu_2\mu_4} + \cr
    &\qquad\left(-4-2(\lambda'-1)+{(\lambda'-1)^2\over24}\right) 
            g_{\mu_1\mu_2}\,g_{\mu_3\mu_4} \;
  \bigg]\; +\cr
  &\left(
   e^{i[-\omega(p_1\!,p_2\!) - \omega(p_1\!,p_3\!)- \omega(p_2\!,p_3\!)]}
   \delta_{i_1}^{j_2}\delta_{i_2}^{j_3}\delta_{i_3}^{j_4}\delta_{i_4}^{j_1}
   \!+
   e^{i[\omega(p_1\!,p_2\!) + \omega(p_1\!,p_3\!)+ \omega(p_2\!,p_3\!)]}
   \delta_{i_1}^{j_4}\delta_{i_2}^{j_1}\delta_{i_3}^{j_2}\delta_{i_4}^{j_3}
   \right)\cr 
    &\quad\bigg[\left(2+(\lambda'-1)+{13(\lambda'-1)^2\over24}\right) 
            g_{\mu_1\mu_4}\,g_{\mu_2\mu_3} +\cr
    &\qquad\left(-4-2(\lambda'-1)+{(\lambda'-1)^2\over24}\right) 
            g_{\mu_1\mu_3}\,g_{\mu_2\mu_4} + \cr
    &\qquad\left(5+{5(\lambda'-1)\over2}+{(\lambda'-1)^2\over24}\right) 
            g_{\mu_1\mu_2}\,g_{\mu_3\mu_4}\;  
  \bigg]\;\;  \Bigg\},\cr
(ii)_{1234}=i\,&\cteI\,N\;\times\cr
\Bigg\{\!
  &\left(
   e^{i[\omega(p_1\!,p_2\!) - \omega(p_1\!,p_3\!)- \omega(p_2\!,p_3\!)]}
   \delta_{i_1}^{j_3}\delta_{i_2}^{j_1}\delta_{i_3}^{j_4}\delta_{i_4}^{j_2}
   \!+
   e^{i[-\omega(p_1\!,p_2\!) + \omega(p_1\!,p_3\!)+ \omega(p_2\!,p_3\!)]}
   \delta_{i_1}^{j_2}\delta_{i_2}^{j_4}\delta_{i_3}^{j_1}\delta_{i_4}^{j_3}
   \right)\cr 
    &\quad\bigg[\left(2+{17(\lambda'-1)\over8}-{(\lambda'-1)^2\over24}\right) 
            g_{\mu_1\mu_4}\,g_{\mu_2\mu_3} +\cr
    &\qquad\left(-{5\over2}-{19(\lambda'-1)\over8}-{(\lambda'-1)^2\over24}\right) 
            g_{\mu_1\mu_3}\,g_{\mu_2\mu_4} + \cr
    &\qquad\left(-{13\over4}-{13(\lambda'-1)\over8}-{13(\lambda'-1)^2\over24}\right) 
            g_{\mu_1\mu_2}\,g_{\mu_3\mu_4} \;
  \bigg]\; +\cr
  &\left(
   e^{i[-\omega(p_1\!,p_2\!) - \omega(p_1\!,p_3\!)- \omega(p_2\!,p_3\!)]}
   \delta_{i_1}^{j_2}\delta_{i_2}^{j_3}\delta_{i_3}^{j_4}\delta_{i_4}^{j_1}
   \!+
   e^{i[\omega(p_1\!,p_2\!) + \omega(p_1\!,p_3\!)+ \omega(p_2\!,p_3\!)]}
   \delta_{i_1}^{j_4}\delta_{i_2}^{j_1}\delta_{i_3}^{j_2}\delta_{i_4}^{j_3}
   \right)\cr 
    &\quad\bigg[\left(-{5\over2}-{19(\lambda'-1)\over8}-{(\lambda'-1)^2\over24}\right) 
            g_{\mu_1\mu_4}\,g_{\mu_2\mu_3} +\cr
    &\qquad\left(2+{17(\lambda'-1)\over8}-{(\lambda'-1)^2\over24}\right) 
            g_{\mu_1\mu_3}\,g_{\mu_2\mu_4} + \cr
    &\qquad\left(-{13\over4}-{13(\lambda'-1)\over8}-{13(\lambda'-1)^2\over24}\right) 
            g_{\mu_1\mu_2}\,g_{\mu_3\mu_4}\;  
  \bigg]\;\;  \Bigg\},\cr
(iii)_{1234}=i\,&\cteI\,N\;\times\cr
\Bigg\{\!
  &\left(
   e^{i[-\omega(p_1\!,p_2\!) - \omega(p_1\!,p_3\!)- \omega(p_2\!,p_3\!)]}
   \delta_{i_1}^{j_2}\delta_{i_2}^{j_3}\delta_{i_3}^{j_4}\delta_{i_4}^{j_1}
   \!+
   e^{i[\omega(p_1\!,p_2\!) + \omega(p_1\!,p_3\!)+ \omega(p_2\!,p_3\!)]}
   \delta_{i_1}^{j_4}\delta_{i_2}^{j_1}\delta_{i_3}^{j_2}\delta_{i_4}^{j_3}
   \right)\cr 
    &\quad\bigg[\left({47\over12}+{5(\lambda'-1)\over2}+{7(\lambda'-1)^2\over12}\right) 
            g_{\mu_1\mu_4}\,g_{\mu_2\mu_3} +\cr
    &\qquad\left({17\over12}-{(\lambda'-1)\over2}+{(\lambda'-1)^2\over12}\right) 
            g_{\mu_1\mu_3}\,g_{\mu_2\mu_4} + \cr
    &\qquad\left({47\over12}+{5(\lambda'-1)\over2}+{7(\lambda'-1)^2\over12}\right) 
            g_{\mu_1\mu_2}\,g_{\mu_3\mu_4} \;
  \bigg]\; \Bigg\},\cr
(iv)_{1234}=&-{i\over24}\,\cteI\,N\;\times\cr
  &\left(
   e^{i[-\omega(p_1\!,p_2\!) - \omega(p_1\!,p_3\!)- \omega(p_2\!,p_3\!)]}
   \delta_{i_1}^{j_2}\delta_{i_2}^{j_3}\delta_{i_3}^{j_4}\delta_{i_4}^{j_1}
   \!+
   e^{i[\omega(p_1\!,p_2\!) + \omega(p_1\!,p_3\!)+ \omega(p_2\!,p_3\!)]}
   \delta_{i_1}^{j_4}\delta_{i_2}^{j_1}\delta_{i_3}^{j_2}\delta_{i_4}^{j_3}
   \right)\cr 
    &\quad\left(
            g_{\mu_1\mu_4}\,g_{\mu_2\mu_3} +
            g_{\mu_1\mu_3}\,g_{\mu_2\mu_4} + 
            g_{\mu_1\mu_2}\,g_{\mu_3\mu_4} \;
  \right).      \cr
}
$$

Summing over permutations, one obtains
$$
\eqalignno{ 
(I)=\;\;\,&(i)_{1234}+(i)_{3214} +(i)_{2134}=\cr
   =i\,&\cteI\,N\;\times\cr
\Bigg\{\!
  &\left(
   e^{i[-\omega(p_1\!,p_2\!) - \omega(p_1\!,p_3\!)+ \omega(p_2\!,p_3\!)]}
   \delta_{i_1}^{j_3}\delta_{i_2}^{j_4}\delta_{i_3}^{j_2}\delta_{i_4}^{j_1}
   \!+
   e^{i[\omega(p_1\!,p_2\!) + \omega(p_1\!,p_3\!)- \omega(p_2\!,p_3\!)]}
   \delta_{i_1}^{j_4}\delta_{i_2}^{j_3}\delta_{i_3}^{j_1}\delta_{i_4}^{j_2}
   \right)\cr 
    &\quad\bigg[\left(7+{7(\lambda'-1)\over2}+{7(\lambda'-1)^2\over12}\right) 
            g_{\mu_1\mu_4}\,g_{\mu_2\mu_3} +\cr
    &\qquad\left(7+{7(\lambda'-1)\over2}+{7(\lambda'-1)^2\over12}\right) 
            g_{\mu_1\mu_3}\,g_{\mu_2\mu_4} + \cr
    &\qquad\left(-8-4(\lambda'-1)+{(\lambda'-1)^2\over12}\right) 
            g_{\mu_1\mu_2}\,g_{\mu_3\mu_4} \;
  \bigg]\; +\cr
  &\left(
   e^{i[-\omega(p_1\!,p_2\!) - \omega(p_1\!,p_3\!)- \omega(p_2\!,p_3\!)]}
   \delta_{i_1}^{j_2}\delta_{i_2}^{j_3}\delta_{i_3}^{j_4}\delta_{i_4}^{j_1}
   \!+
   e^{i[\omega(p_1\!,p_2\!) + \omega(p_1\!,p_3\!)+ \omega(p_2\!,p_3\!)]}
   \delta_{i_1}^{j_4}\delta_{i_2}^{j_1}\delta_{i_3}^{j_2}\delta_{i_4}^{j_3}
   \right)\cr 
    &\quad\bigg[\left(7+{7(\lambda'-1)\over2}+{7(\lambda'-1)^2\over12}\right) 
            g_{\mu_1\mu_4}\,g_{\mu_2\mu_3} +\cr
    &\qquad\left(-8-4(\lambda'-1)+{(\lambda'-1)^2\over12}\right) 
            g_{\mu_1\mu_3}\,g_{\mu_2\mu_4} + \cr
    &\qquad\left(7+{7(\lambda'-1)\over2}+{7(\lambda'-1)^2\over12}\right) 
            g_{\mu_1\mu_2}\,g_{\mu_3\mu_4}\;  
  \bigg]\; +\cr
  &\left(
   e^{i[\omega(p_1\!,p_2\!) - \omega(p_1\!,p_3\!)- \omega(p_2\!,p_3\!)]}
   \delta_{i_1}^{j_3}\delta_{i_2}^{j_1}\delta_{i_3}^{j_4}\delta_{i_4}^{j_2}
   \!+
   e^{i[-\omega(p_1\!,p_2\!) + \omega(p_1\!,p_3\!)+ \omega(p_2\!,p_3\!)]}
   \delta_{i_1}^{j_2}\delta_{i_2}^{j_4}\delta_{i_3}^{j_1}\delta_{i_4}^{j_3}
   \right)\cr 
    &\quad\bigg[\left(-8-4(\lambda'-1)+{(\lambda'-1)^2\over12}\right) 
            g_{\mu_1\mu_4}\,g_{\mu_2\mu_3} +\cr
    &\qquad\left(7+{7(\lambda'-1)\over2}+{7(\lambda'-1)^2\over12}\right) 
            g_{\mu_1\mu_3}\,g_{\mu_2\mu_4} + \cr
    &\qquad\left(7+{7(\lambda'-1)\over2}+{7(\lambda'-1)^2\over12}\right) 
            g_{\mu_1\mu_2}\,g_{\mu_3\mu_4}\;  
  \bigg]\;\;\Bigg\},\cr
(II)=\;\;\,&(ii)_{1234}+(ii)_{1324} +(ii)_{1423}+(ii)_{2314}
         +(ii)_{2413}+(ii)_{3412}=\cr
=i\,&\cteI\,N\;\times\cr
\Bigg\{\!
  &\left(
   e^{i[\omega(p_1\!,p_2\!) - \omega(p_1\!,p_3\!)- \omega(p_2\!,p_3\!)]}
   \delta_{i_1}^{j_3}\delta_{i_2}^{j_1}\delta_{i_3}^{j_4}\delta_{i_4}^{j_2}
   \!+
   e^{i[-\omega(p_1\!,p_2\!) + \omega(p_1\!,p_3\!)+ \omega(p_2\!,p_3\!)]}
   \delta_{i_1}^{j_2}\delta_{i_2}^{j_4}\delta_{i_3}^{j_1}\delta_{i_4}^{j_3}
   \right)\cr 
    &\quad\bigg[\left(8+{17(\lambda'-1)\over2}-{(\lambda'-1)^2\over6}\right) 
            g_{\mu_1\mu_4}\,g_{\mu_2\mu_3} +\cr
    &\qquad\left(-{23\over2}-{8(\lambda'-1)}-{7(\lambda'-1)^2\over6}\right) 
            g_{\mu_1\mu_3}\,g_{\mu_2\mu_4} + \cr
    &\qquad\left(-{23\over2}-{8(\lambda'-1)}-{7(\lambda'-1)^2\over6}\right) 
            g_{\mu_1\mu_2}\,g_{\mu_3\mu_4} \;
  \bigg]\; +\cr
  &\left(
   e^{i[-\omega(p_1\!,p_2\!) - \omega(p_1\!,p_3\!)- \omega(p_2\!,p_3\!)]}
   \delta_{i_1}^{j_2}\delta_{i_2}^{j_3}\delta_{i_3}^{j_4}\delta_{i_4}^{j_1}
   \!+
   e^{i[\omega(p_1\!,p_2\!) + \omega(p_1\!,p_3\!)+ \omega(p_2\!,p_3\!)]}
   \delta_{i_1}^{j_4}\delta_{i_2}^{j_1}\delta_{i_3}^{j_2}\delta_{i_4}^{j_3}
   \right)\cr 
    &\quad\bigg[\left(-{23\over2}-{8(\lambda'-1)}-{7(\lambda'-1)^2\over6}\right) 
            g_{\mu_1\mu_4}\,g_{\mu_2\mu_3} +\cr
    &\qquad\left(8+{17(\lambda'-1)\over2}-{(\lambda'-1)^2\over6}\right) 
            g_{\mu_1\mu_3}\,g_{\mu_2\mu_4} + \cr
    &\qquad\left(-{23\over2}-{8(\lambda'-1)}-{7(\lambda'-1)^2\over6}\right) 
            g_{\mu_1\mu_2}\,g_{\mu_3\mu_4}\;  
  \bigg]\;  +\cr  
  &\left(
   e^{i[-\omega(p_1\!,p_2\!) - \omega(p_1\!,p_3\!)+ \omega(p_2\!,p_3\!)]}
   \delta_{i_1}^{j_3}\delta_{i_2}^{j_4}\delta_{i_3}^{j_2}\delta_{i_4}^{j_1}
   \!+
   e^{i[\omega(p_1\!,p_2\!) + \omega(p_1\!,p_3\!)- \omega(p_2\!,p_3\!)]}
   \delta_{i_1}^{j_4}\delta_{i_2}^{j_3}\delta_{i_3}^{j_1}\delta_{i_4}^{j_2}
   \right)\cr 
    &\quad\bigg[\left(-{23\over2}-{8(\lambda'-1)}-{7(\lambda'-1)^2\over6}\right) 
            g_{\mu_1\mu_4}\,g_{\mu_2\mu_3} +\cr
    &\qquad\left(-{23\over2}-{8(\lambda'-1)}-{7(\lambda'-1)^2\over6}\right) 
            g_{\mu_1\mu_3}\,g_{\mu_2\mu_4} + \cr
    &\qquad\left(8+{17(\lambda'-1)\over2}-{(\lambda'-1)^2\over6}\right) 
            g_{\mu_1\mu_2}\,g_{\mu_3\mu_4}\;  
  \bigg]\;\;  \Bigg\},\cr
(III)=\;\;\,&(iii)_{1234}+(iii)_{1324} +(iii)_{1243}=\cr
     =i\,&\cteI\,N\;\times\cr
\Bigg\{\!
  &\left(
   e^{i[-\omega(p_1\!,p_2\!) - \omega(p_1\!,p_3\!)- \omega(p_2\!,p_3\!)]}
   \delta_{i_1}^{j_2}\delta_{i_2}^{j_3}\delta_{i_3}^{j_4}\delta_{i_4}^{j_1}
   \!+
   e^{i[\omega(p_1\!,p_2\!) + \omega(p_1\!,p_3\!)+ \omega(p_2\!,p_3\!)]}
   \delta_{i_1}^{j_4}\delta_{i_2}^{j_1}\delta_{i_3}^{j_2}\delta_{i_4}^{j_3}
   \right)\cr 
    &\quad\bigg[\left({47\over12}+{5(\lambda'-1)\over2}+{7(\lambda'-1)^2\over12}\right) 
            g_{\mu_1\mu_4}\,g_{\mu_2\mu_3} +\cr
    &\qquad\left({17\over12}-{(\lambda'-1)\over2}+{(\lambda'-1)^2\over12}\right) 
            g_{\mu_1\mu_3}\,g_{\mu_2\mu_4} + \cr
    &\qquad\left({47\over12}+{5(\lambda'-1)\over2}+{7(\lambda'-1)^2\over12}\right) 
            g_{\mu_1\mu_2}\,g_{\mu_3\mu_4} \;
  \bigg]\; +\cr
  &\left(
   e^{i[-\omega(p_1\!,p_2\!) - \omega(p_1\!,p_3\!)+ \omega(p_2\!,p_3\!)]}
   \delta_{i_1}^{j_3}\delta_{i_2}^{j_4}\delta_{i_3}^{j_2}\delta_{i_4}^{j_1}
   \!+
   e^{i[\omega(p_1\!,p_2\!) + \omega(p_1\!,p_3\!)- \omega(p_2\!,p_3\!)]}
   \delta_{i_1}^{j_4}\delta_{i_2}^{j_3}\delta_{i_3}^{j_1}\delta_{i_4}^{j_2}
   \right)\cr 
    &\quad\bigg[\left({47\over12}+{5(\lambda'-1)\over2}+{7(\lambda'-1)^2\over12}\right) 
            g_{\mu_1\mu_4}\,g_{\mu_2\mu_3} +\cr
    &\qquad\left({47\over12}+{5(\lambda'-1)\over2}+{7(\lambda'-1)^2\over12}\right) 
            g_{\mu_1\mu_3}\,g_{\mu_2\mu_4} + \cr
    &\qquad\left({17\over12}-{(\lambda'-1)\over2}+{(\lambda'-1)^2\over12}\right) 
            g_{\mu_1\mu_2}\,g_{\mu_3\mu_4} \;
  \bigg]\; +\cr
  &\left(
   e^{i[\omega(p_1\!,p_2\!) - \omega(p_1\!,p_3\!)- \omega(p_2\!,p_3\!)]}
   \delta_{i_1}^{j_3}\delta_{i_2}^{j_1}\delta_{i_3}^{j_4}\delta_{i_4}^{j_2}
   \!+
   e^{i[-\omega(p_1\!,p_2\!) + \omega(p_1\!,p_3\!)+ \omega(p_2\!,p_3\!)]}
   \delta_{i_1}^{j_2}\delta_{i_2}^{j_4}\delta_{i_3}^{j_1}\delta_{i_4}^{j_3}
   \right)\cr 
    &\quad\bigg[\left({17\over12}-{(\lambda'-1)\over2}+{(\lambda'-1)^2\over12}\right) 
            g_{\mu_1\mu_4}\,g_{\mu_2\mu_3} +\cr
    &\qquad\left({47\over12}+{5(\lambda'-1)\over2}+{7(\lambda'-1)^2\over12}\right) 
            g_{\mu_1\mu_3}\,g_{\mu_2\mu_4} + \cr
    &\qquad\left({47\over12}+{5(\lambda'-1)\over2}+{7(\lambda'-1)^2\over12}\right) 
            g_{\mu_1\mu_2}\,g_{\mu_3\mu_4} \;
  \bigg]\; \;\Bigg\},\cr
(IV)=\;\;\,&(iv)_{1234}+(iv)_{1324} +(iv)_{1243}+
            (iv)_{1432}+(iv)_{1342} +(iv)_{1423}=\cr
    =&-{i\over12}\,\cteI\,N\;\times\cr
  &\Bigg(
   e^{i[-\omega(p_1\!,p_2\!) - \omega(p_1\!,p_3\!)- \omega(p_2\!,p_3\!)]}
   \delta_{i_1}^{j_2}\delta_{i_2}^{j_3}\delta_{i_3}^{j_4}\delta_{i_4}^{j_1}
   \!+
   e^{i[\omega(p_1\!,p_2\!) + \omega(p_1\!,p_3\!)+ \omega(p_2\!,p_3\!)]}
   \delta_{i_1}^{j_4}\delta_{i_2}^{j_1}\delta_{i_3}^{j_2}\delta_{i_4}^{j_3}
   \!+\cr
  &\;\;
   e^{i[-\omega(p_1\!,p_2\!) - \omega(p_1\!,p_3\!)+ \omega(p_2\!,p_3\!)]}
   \delta_{i_1}^{j_3}\delta_{i_2}^{j_4}\delta_{i_3}^{j_2}\delta_{i_4}^{j_1}
   \!+
   e^{i[\omega(p_1\!,p_2\!) + \omega(p_1\!,p_3\!)- \omega(p_2\!,p_3\!)]}
   \delta_{i_1}^{j_4}\delta_{i_2}^{j_3}\delta_{i_3}^{j_1}\delta_{i_4}^{j_2}
   \!+\cr
  &\;\;
    e^{i[\omega(p_1\!,p_2\!) - \omega(p_1\!,p_3\!)- \omega(p_2\!,p_3\!)]}
   \delta_{i_1}^{j_3}\delta_{i_2}^{j_1}\delta_{i_3}^{j_4}\delta_{i_4}^{j_2}
   \!+
   e^{i[-\omega(p_1\!,p_2\!) + \omega(p_1\!,p_3\!)+ \omega(p_2\!,p_3\!)]}
   \delta_{i_1}^{j_2}\delta_{i_2}^{j_4}\delta_{i_3}^{j_1}\delta_{i_4}^{j_3}  
   \Bigg)\cr 
    &\quad\left(
            g_{\mu_1\mu_4}\,g_{\mu_2\mu_3} +
            g_{\mu_1\mu_3}\,g_{\mu_2\mu_4} + 
            g_{\mu_1\mu_2}\,g_{\mu_3\mu_4} \;
  \right).     \cr
}
$$
The sum 
$$
(I)\,+\,(II)\,+\,(III)\,+\,(IV)
$$
is the UV divergent part of the 1PI 4-point gauge function in the MS scheme. 
Note that (I), (II), (III) and (IV) are not proportional to the tree-level 
4-point vertex.

\bigskip

\section{Appendix D. Ghost functions}
The diagrams needed to compute the ghost self-energy and the ghost-boson
vertex function are shown in Figure 6.

\midinsert
{\settabs 3\columns \def\graphwidth{1.6in}   
\eightpoint
\+\hfill\epsfxsize=\graphwidth\epsffile{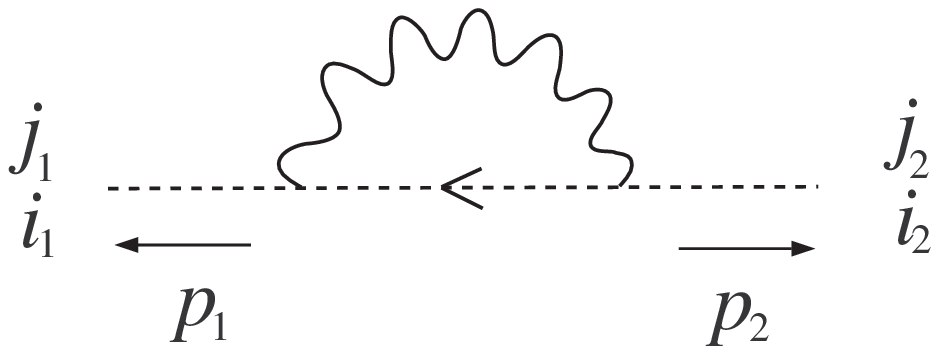}\hfill
	&\hfill\epsfxsize=\graphwidth\epsffile{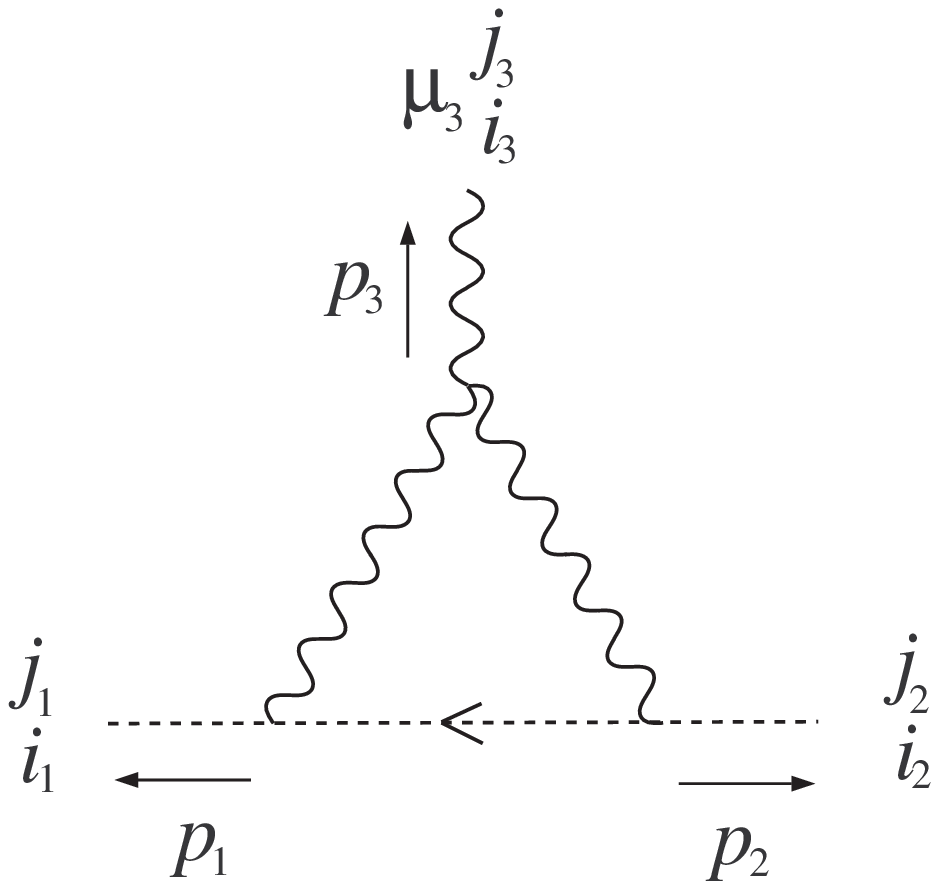}\hfill
	&\hfill\epsfxsize=\graphwidth\epsffile{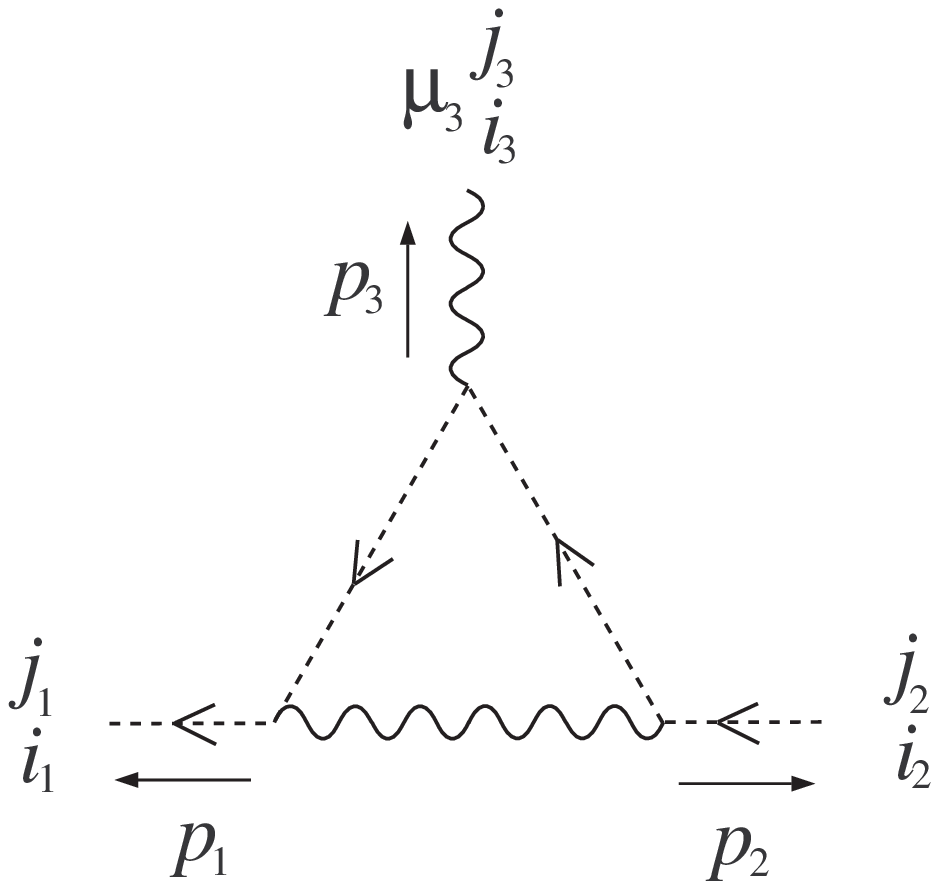}\hfill &\cr
\+\hfill$(i)$\hfill
	&\hfill $(ii)$\hfill
	& \hfill $(iii)$\hfill &\cr
}

\vskip 12pt
\narrower\noindent {\bf Figure 6:}
{\eightrm 1PI UV divergent Feynman digrams for the ghost  functions.}
\vskip 0.4cm
\endinsert
From them, we have obtained the following results
$$
\eqalignno{
(i)=&
  -i \cteI\;g^2\;\Bigl(1-{\lambda'-1\over2}\Bigr)\,N\,
  \delta_{i_1}^{j_2} \delta_{i_2}^{j_1}\; p_1^{\,2}, \cr
(ii)=&3{i\over4}\cteI\,g^2\;\bigl(1+(\lambda'-1)\bigr)\cr
  &N\,\left[e^{-i\omega(p_1,p_2)}
            \delta_{i_1}^{j_2}\,\delta_{i_2}^{j_3}\,\delta_{i_3}^{j_1}-
            e^{i\omega(p_1,p_2)},
            \delta_{i_1}^{j_3}\,\delta_{i_2}^{j_1}\,\delta_{i_3}^{j_2}
      \right]\;p_{1\,\mu_3}, &\numeq\cr
(iii)=&{i\over4}\cteI\,g^2\;\bigl(1+(\lambda'-1)\bigr)\cr
  &N\,\left[e^{-i\omega(p_1,p_2)}
            \delta_{i_1}^{j_2}\,\delta_{i_2}^{j_3}\,\delta_{i_3}^{j_1}-
            e^{i\omega(p_1,p_2)}
            \delta_{i_1}^{j_3}\,\delta_{i_2}^{j_1}\,\delta_{i_3}^{j_2}
      \right]\;p_{1\,\mu_3}. \cr
}
$$

\bigskip

\section{Appendix E. Functions with external fields}

The diagrams which contribute at the one-loop level are shown in Figure 7.
\midinsert
{\settabs 6\columns \def\graphwidth{1.4in} 
\eightpoint

\+&\hfil$\vcenter{$\,$\epsfxsize=1.2in\epsffile{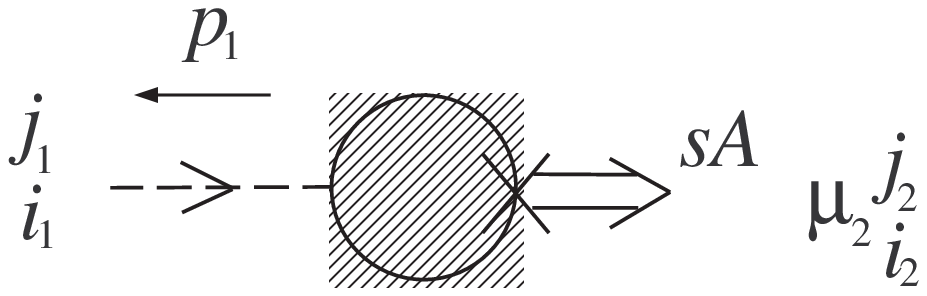}}$\hfil
      & $\qquad\vcenter{=}$
      & \hfil$\vcenter{\epsfxsize=\graphwidth\epsffile{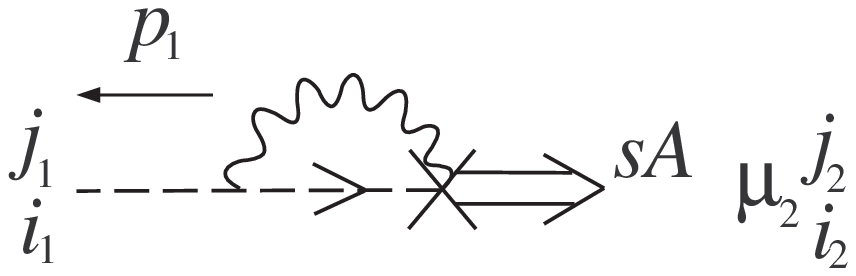}}$\hfil
                     \cr
\+&&&\hfil$\;\;(i)$& \cr
\+&\hfil$\vcenter{\epsfxsize=\graphwidth\epsffile{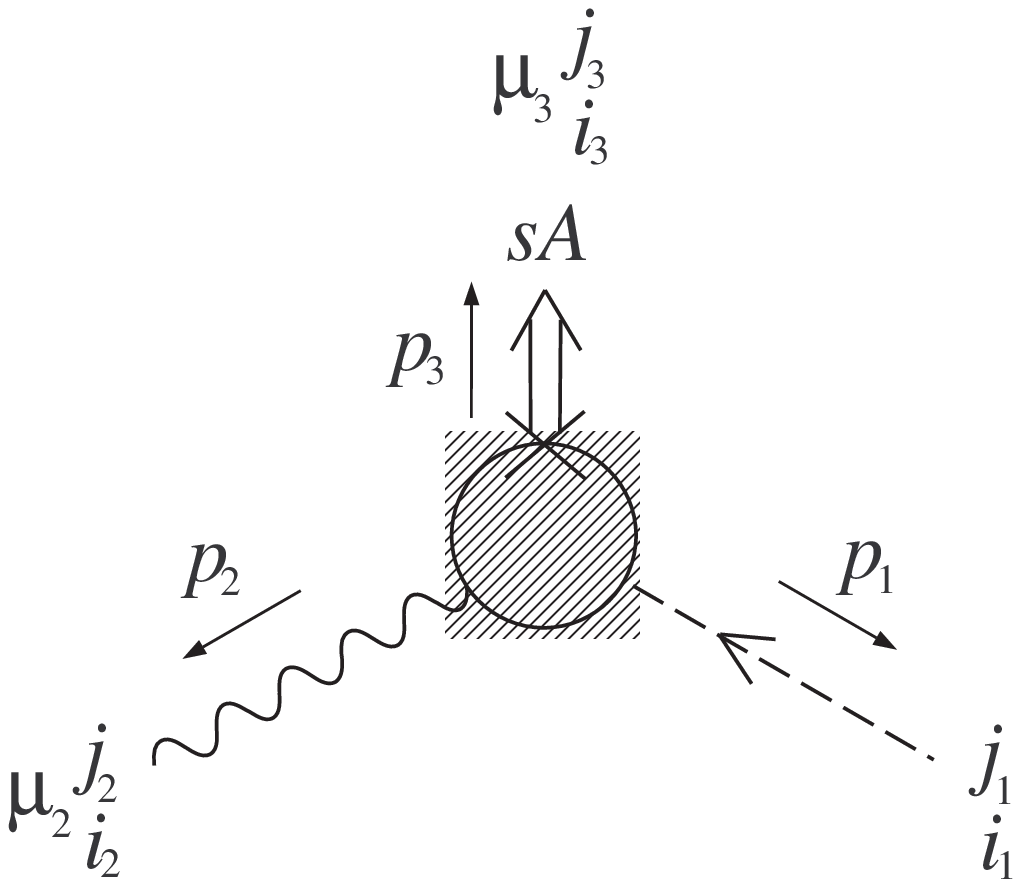}}$\hfil
      & $\qquad\vcenter{=}$
      & \hfil$\vcenter{\epsfxsize=0.9in\epsffile{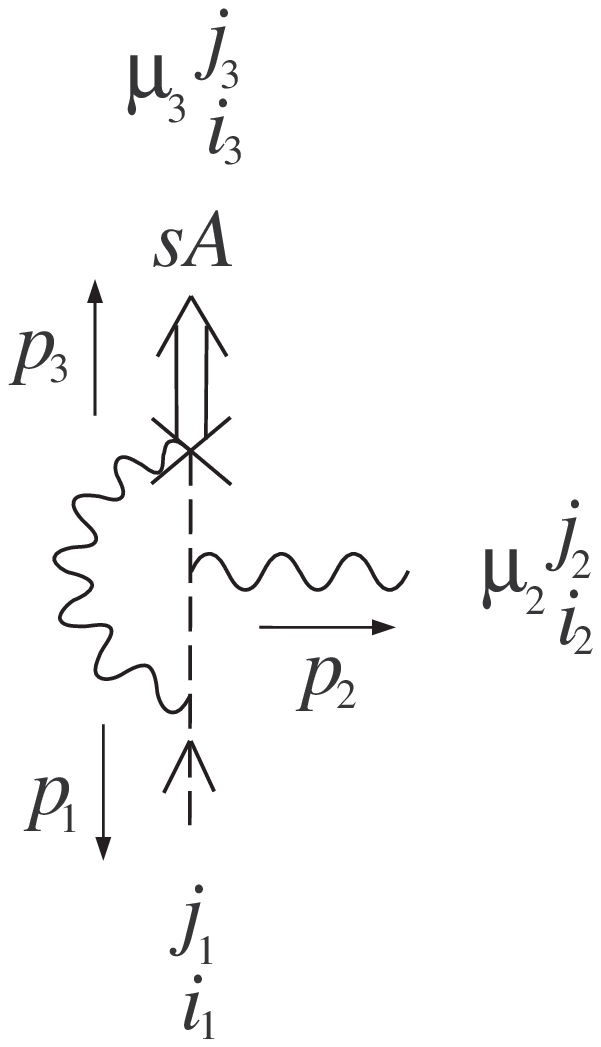}}$\hfil
      & $+
         \quad
         \vcenter{\epsfxsize=\graphwidth\epsffile{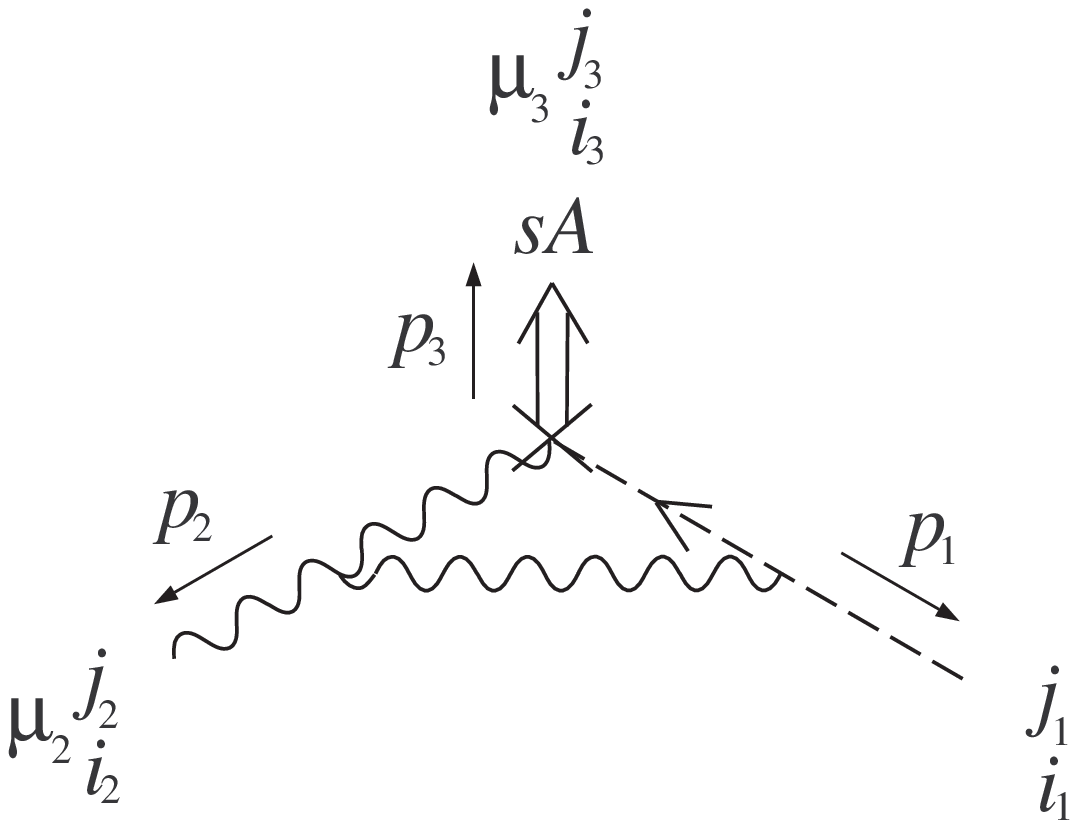}}$
                     \cr
\+&&&\hfil$\;\;(ii)$&&$\;\;(iii)$& \cr
\+&\hfil$\vcenter{$\,$\epsfxsize=1.2in\epsffile{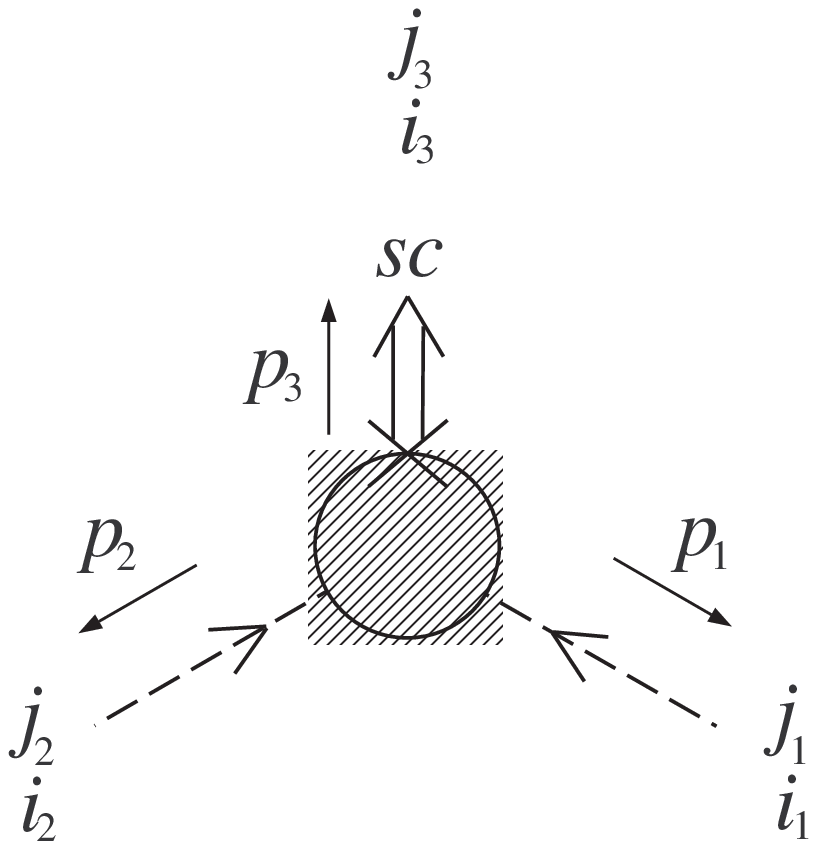}}$\hfil
      & $\qquad\vcenter{=}$
      & \hfil$\vcenter{\epsfxsize=\graphwidth\epsffile{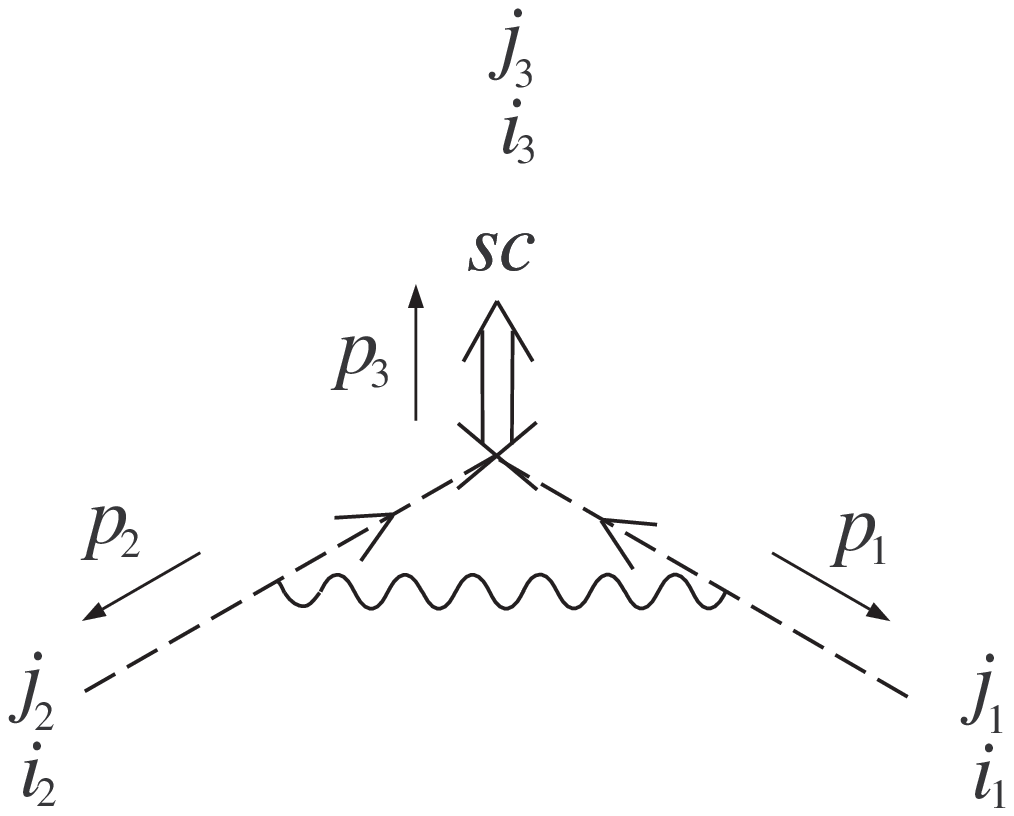}}$\hfil
                     \cr
\+&&&\hfil$\;\;(iv)$& \cr

}
\vskip 12pt
\narrower\noindent {\bf Figure 7:}
{\eightrm 1PI UV divergent digrams involving the external fields.}

\vskip 0.4cm
\endinsert
The results are:
$$
\eqalignno{
(i)=
  &-i\Bigl({1\over(4\pi)^2\varepsilon}\Bigr)
   \,g^2\,\TR\;\Bigl(1 -{\lambda'-1\over2}\Bigr) 
  N\;\delta_{i_1}^{j_2}\,\delta_{i_2}^{j_1}\;p_{1\,\mu_2}, \cr
(ii)=
&-{i\over4}\cteI\,\,g^2\,\TR\;\bigl(1+(\lambda'-1)\bigr)\;\times\cr
 &N\,\left[e^{-i\omega(p_1,p_2)}
            \delta_{i_1}^{j_2}\,\delta_{i_2}^{j_3}\,\delta_{i_3}^{j_1}-
            e^{i\omega(p_1,p_2)}
            \delta_{i_1}^{j_3}\,\delta_{i_2}^{j_1}\,\delta_{i_3}^{j_2}
      \right]\;g_{\mu_2\mu_3},      \cr
(iii)=
&-{3i\over4}\cteI\,\,g^2\,\TR\;\bigl(1+(\lambda'-1)\bigr)\;\times\cr
 &N\,\left[e^{-i\omega(p_1,p_2)}
            \delta_{i_1}^{j_2}\,\delta_{i_2}^{j_3}\,\delta_{i_3}^{j_1}-
            e^{i\omega(p_1,p_2)}
            \delta_{i_1}^{j_3}\,\delta_{i_2}^{j_1}\,\delta_{i_3}^{j_2}
      \right]\;g_{\mu_2\mu_3}, \cr
(iv)=
&i\Bigl({1\over(4\pi)^2\varepsilon}\Bigr)\,g^2\,\TR\;
   \bigl(1+(\lambda'-1)\bigr)
 N\,\left[e^{-i\omega(p_1,p_2)}
            \delta_{i_1}^{j_2}\,\delta_{i_2}^{j_3}\,\delta_{i_3}^{j_1}-
            e^{i\omega(p_1,p_2)}
            \delta_{i_1}^{j_3}\,\delta_{i_2}^{j_1}\,\delta_{i_3}^{j_2}
      \right].      \cr
}
$$

\bigskip

\section{Acknowledgments} 
This work has been partially supported by CICyT under grant PB98-0842.

\section{References}

\frenchspacing

\refno\Tony.
A. Gonz\'alez-Arroyo and M. Okawa, Phys. Lett. {\bf B120} (1983) 174;
T. Eguchi and R. Nakayama, Phys. Lett. {\bf B122} (1983) 59; 
A. Gonz\'alez-Arroyo and M. Okawa, Phys. Rev. {\bf D27} (1983) 2397;
A. Gonz\'alez-Arroyo and C. P. Korthals-Altes, Phys. Lett. {\bf B131} (1983) 
396.

\refno\Makeenko.
Y. Makeenko, {\it Reduced Models and Noncommutative Gauge Theories},
{\tt hep-th/000928}.

\refno\Doplicher.
S. Doplicher, K. Fredenhagen and J.E. Roberts, Phys. Letts. B331 (1994) 39;
S. Doplicher, K. Fredenhagen and J.E. Roberts, Commun. Math. Phys. 172 (1995)
187.

\refno\CDS.
A. Connes, M.R. Douglas and A. Schwarz, J. High Energy. Phys. 02 (1998) 003;
{\tt hep-th/9711162}.

\refno\Mtheory.
M.R. Douglas and
C. Hull, J. High Energy Phys. 9802 (1998) 008; H. Li, {\it Comments on 
Supersymmetric Yang-Mills Theory on a Noncommutative Torus}, 
{\tt hep-th/9802052};
R. Casalbuoni, Phys. Lett. B431 (1998)69;
A. Schwarz, Nucl. Phys. B534 (1998)720;
T. Kawano and K. Okuyama, Phys. Lett. B433 (1998)29; 
Y.-K Cheung and M. Krog, Nucl. Phys. B528(1998)185
F. Ardalan, H. Arfaei 
and M.M. Sheikh-Jabbari, {\tt hep-th/9803067}; 
G. Landi, F. Lizzi and R.J. Szabo, Commun.Math.Phys. {\bf 206} (1999) 603; 
B. Moriaru and B. Zumino,
{\tt hep-th/9807198}; C. Hofman and E. Verlinde, Nucl. Phys. {\bf B547} (1999) 
157.

\refno\Branes.
C.-S. Chu and P.-H. Ho, Nucl. Phys. {\bf B550} (1999) 151;
F. Ardalan, H. Arfei and M.M. Sheikh-Jabbari, J. High Energy Phys. {\bf 02} 
 (1999); 
V. Schomerus, J. High Energy Phys. {\bf 06} (1999) 030;
L. Cornalba, {\it D-brane Physics and Noncommutative Yang-Mills Theory},
{\tt hep-th/9909091};
N. Seiberg and E. Witten, J. High Energy Phys. {\bf 09} (1999) 032;
O. Andreev and H. Dorn, Nucl. Phys. {\bf B583} (2000) 145;
A. Bilal, C.-S. Chu and R. Russo, {\bf B582} (2000) 65;
Y. Kien and S. Lee, Nucl. Phys. {\bf B586} (2000) 303;
H. Liu and J. Michelson, Phys. Rev. {\bf D62} (2000) 066003.

\refno\Filk.
T. Filk, Phys. Lett. B376 (1996)53.

\refno\Continuum.
A very incomplete list of references reads: 
N. Nekrasov and A.S. Schwarz, Comm. Math. Phys. {\bf 198} (1998) 689;
J.C. V\'{a}rilly and J.M. Grac\'{\i}a-Bondia, Int.J.Mod.Phys. {\bf A14} 
(1999) 1305;
M. Chaichian, A. Demichev and P. Pre{\v s}najder, Nucl.Phys. {\bf B567} 
(2000) 360;
P. Kosi\'nski, J. Lukierski and P. Ma\'slanka, Phys.Rev. {\bf D62} (2000) 
025004;
T. Krajewski and R. Wulkenhaar, Int.J.Mod.Phys. {\bf A15} (2000) 1011;
C. P. Mart\'{\i}n and D. S\'anchez-Ruiz, Phys. Rev. Lett. {\bf 83} (1999) 476.
M. Sheikh-Jabbari, J. High Energy Phys. {\bf 06} (1999) 015.
S. Cho, R. Hinderting, J. Madore and H. Steinacker, 
Int. J. Mod. Phys. {\bf D9} (2000) 161; I. Y. Aref'eva, D.M. Belov and 
A.S  Koshelev, Phys.Lett. {\bf B476} (2000) 431; 
C.-S Chu, Nucl.Phys. {\bf B580} (2000) 352; 
B. A. Campbell and K. Kaminsky, Nucl.Phys. {\bf B581} (2000) 240; 
F. Zamora, J. High Energy Phys. {\bf  0005} (2000) 002; 
W. Fischler, E. Gorbatov, A. Kashani-Poor, R. McNees, S. Paban and 
P. Pouliot, J. High Energy Phys. {\bf 0006} (2000) 032;
G. Arcioni, J.L.F. Barbon, Joaquim Gomis and M.A. Vazquez-Mozo, 
J. High Energy Phys. {\bf  0006} (2000) 038;
K. Furuta and T. Inami, Mod. Phys. Lett. {\bf A15} (2000) 997; 
G.-H. Chen and Y.-S Wu, {\it One Loop Shift in Noncommutative Chern-Simons
Coupling}, {\tt hep-th/0006114};
J. Gomis, T. Mehen and M. B. Wise, J. High Energy Phys. {\bf  0008}
 (2000) 029;  Phys.Lett. {\bf B491} (2000) 345; 
L. Alvarez-Gaume and J.L.F. Barbon, {\it Nonlinear Vacuum Phenomena in Noncommutative QED}, {\tt hep-th/0006209 };
K. Landsteiner, E. Lopez and M.G.H. Tytgat, J. High Energy Phys. 
 {\bf 0009} (2000) 027;
J. W. Moffat, Phys.Lett. {\bf B491} (2000) 345;
E. F. Moreno and  F.A. Schaposnik, {\it Wess-Zumino-Witten and
Fermion Models in Noncommutative Space}, {\tt hep-th/0008118}; 
I.F. Riad and M. M. Sheikh-Jabbari, J. High Energy Phys. {\bf  0008} 
(2000) 045; S. Elitzur, B. Pioline and E. Rabinovici, J. High Energy Phys. 
{\bf 0010} (2000) 011; 
D.J. Gross, A. Hashimoto and N. Itzhaki, {\it Observables of Noncommutative 
Gauge Theories}, {\tt hep-th/0008075}
W.-H. Huang, {\it Two Loop Effective Potential in 
Noncommutative Scalar Field Theory}, {\tt hep-th/0009067 };
N. Grandi and  G.A. Silva, {\it Chern-Simons action in Noncommutative Space},
{\tt hep-th/0010113}; H. Arfaei and  M.H. Yavartanoo, 
{\it Phenomenological Consequences of Noncommuative QED},{\tt hep-th/0010244}; 
T. Mehen and M. B. Wise, {\it Generalized *-Products, Wilson  lines and the
solution of the Seiberg-Witten Equations}, {\tt hep-th/0010204};
W.-H Huang, {\it Finite Temperature Casimir Effect and the Radius 
Stabilization of the Noncommutative Torus}, {\tt hep-th/0011037};
M. Pernici, A. Santambrogio and D. Zanon, {\it The one-loop effective action of noncommutative ${\cal N}=4$ super Yang-Mills is gauge invariant}, 
{\tt  hep-th/0011140};  D. Zanon, {\it Noncommutative ${\cal N}=1,2$ super $U(N)$ Yang-Mills: UV/IR mixing and effective action results at one loop}, 
{\tt hep-th/0012009 }.

\refno\Lattice.
J. Ambjorn, Y.M. Makeenko, J. Nishimura and  R.J. Szabo, 
Phys. Lett. {\bf  B480} (2000) 399; 
J. High Energy Phys. {\bf 0005} (2000) 023. 

\refno\Phenomenon.
Irina Mocioiu, Maxim Pospelov and Radu Roiban, Phys. Lett. {\bf B489} (2000) 
390;
J.L.  Hewett, F.J. Petriello and T.G. Rizzo, {\it Signals for Non-Commutative
Interactions at Linear Colliders}, {\tt hep-ph/0010354};
C.-S. Chu, B.R. Greene and G. Shiu, {\it Remarks on Inflation and 
Noncommutative Geometry}, {\tt hep-th/0011241};
Prakash Mathews, {\it Compton scattering in Noncommutative Space-Time at the 
NLC}, {\tt hep-ph/0011332}.
 
\refno\MRS.
S. Minwalla, M. Van Raamsdonk and N. Seiberg, {\it Noncommutative Perturbative Dynamics}, {\tt hep-th/9912027}. 

\refno\UVIR.
M. Hayakawa, Phys. Lett.  {\bf B478} (2000) 394; 
A. Matusis, L. Susskind and  N. Toumbas, 
{\it The IR/UV Connection in the Non-Commutative Gauge Theories},
{\tt hep-th/0002075};
J. Gomis, K. Landsteiner, E. Lopez, Phys.Rev. {\bf D62} (2000) 105006;
J. Gomis, T. Mehen, M. B. Wise, J. High Energy Phys. {\bf  0008} (2000) 029; 
C.P. Mart\'{\i}n and F. Ruiz Ruiz, {\it Paramagnetic dominance, the sign of 
the beta function and UV/IR mixing in non-commutative U(1)}, 
{\tt hep-th/0007131}. 

\refno\NCBPHZ.
I. Chepelev and R. Roiban, J. High Energy Phys. {\bf 0005} (2000) 037; 
{\it Convergence Theorem for Non-commutative Feynman Graphs and Renormalization}, {\tt hep-th/0008090}.

\refno\BPHZ.
N.N. Bogoliubov and O.S. Parasiuk, Acta Math. {\bf 97} (1957) 227; K. Hepp, 
Comm. Math, Phys. {\bf 2} (1966) 301; W. Zimmerman {\it Local Operator Products and Renormalization in Quantum Field Theory}, 1970 Brandeis University Summer
Institute in Thoeretical Physics, The M.I.T. Press, Edited by S. Deser, M. Grisaru and H. Pendleton.

\refno\Susy.
H.O. Girotti, M. Gomes, V.O. Rivelles  and A.J. da Silva, 
Nucl.Phys. {\bf B587} (2000) 299; 
A.A. Bichl, J.M. Grimstrup, H. Grosse, L. Popp, M. Schweda and R. Wulkenhaar,
{\it The Superfield Formalism Applied to the Noncommutative 
Wess-Zumino Model}, {\tt hep-th/0007050}.

\refno\Wilson.
K. G. Wilson, Phys. Rev. {\bf B4} (1971) 3174; Phys. Rev. {\bf B4} (1971) 3184;
J. Polchinski, Nuc. Phys {\bf B231} (1984) 269.

\refno\Firstorder.
S.S. Gubser and  S. L. Sondhi, 
{\it Phase structure of non-commutative scalar field theories}, 
{\tt hep-th/0006119}. 

\refno\Nonren.
I. Ya. Aref'eva, D. M. Belov and  A. S.Koshelev,
{\it A Note on UV/IR for Noncommutative Complex Scalar Field},
{\tt  hep-th/0003176}.

\refno\QAP.
J.H. Lowenstein, Phys. Rev. {\bf D4} (1971) 2281; Comm. Math. Phys. {\bf 24} 
(1971) 1; Y.M.P. Lam,  Phys. Rev. {\bf D6} (1972) 2145; 
Phys. Rev. {\bf D7} (1973) 2943; T.E. Clark and J.H. Lowenstein, Nucl. Phys. 
{\bf B113} (1976) 109.

\refno\Algren.
O. Piguet and S.P. Sorella, {\it Algebraic Renormalization}, Lectures notes
in Physics, Monographs, Springer-Verlag 1995;
J. Gomis, J. Paris and S. Samuel, Phys. Rep {\bf 259} (1995) 1;
G. Barnich, F. Brandt and M. Henneaux, Phys. Rep. {\bf 338} (2000) 439.

\refno\Unitarity.
J. Gomis and T. Mehen, Nucl. Phys. {\bf B591} (2000) 265;
O. Aharony, J. Gomis and T. Mehen, J. High Energy Phys. {\bf 0009} (2000) 023.

\refno\Books.
J.M. Gracia-Bond\'{\i}a, J.C. V\'arilly and H. Figueroa, {\it Elements of 
Noncommutative Geometry}, Birkhauser, 2000; G. Landi, {\it An Introduction to
Noncommutative Spaces and their Geometries}, Springer, Lecture Notes in 
Physics 51, Springer Verlag, 1997; 
J. Madore, {\it An Introduction to Noncommutative Geometry and its applications}, LMS, Cambridge University Press, 1995; A. Connes, {\it Noncommutative Geometry}, 
Academic Press, 1994.

\refno\Hooft.
G. 't Hooft, Nucl. Phys. {\bf B72} (1974) 461.

\refno\BM.
P. Breitenlohner and Maison, Comm. Math. Phys. {\bf 52} (1977) 11, 
{\it ibid.} {\bf 52} (1977) 39, {\bf 52} (1977) 55.

\refno\NCUN.
A. Armoni, {\it Comments on Perturbative Dynamics of Non-Commutative Yang-Mills
Theory}, {\tt hep-th/0005208}; L. Bonora and M. Salizzoni, {\it Renormalization
of noncommutative $U(N)$ gauge theories}, {\tt hep-th/0011088}.

\bye